\newif\ifdraft
\drafttrue  % SET DRAFT
\ifdraft
\documentclass[onecolumn]{IEEEtran}
\else
\documentclass[journal]{IEEEtran}
\fi

\usepackage[utf8]{inputenc}

\usepackage{amsmath,amssymb}
\usepackage{algorithm2e}
\usepackage{color}
\usepackage{graphicx}
\usepackage{nameref,hyperref}
\usepackage{url}
\usepackage{subfig}

\newcommand{\changes}{\color{black}}

\begin{document}

\title{Principal Patterns on Graphs: Discovering Coherent Structures in Datasets}

\author{
    \IEEEauthorblockN{Kirell~Benzi, Benjamin~Ricaud and~Pierre~Vandergheynst} \\
    \IEEEauthorblockA{Laboratoire de Traitement des Signaux 2, \\ École Polytechnique Fédérale de Lausanne, Lausanne, Vaud, Switzerland
    \\\{first.last\}@epfl.ch}
}

\maketitle

\begin{abstract}
Graphs are now ubiquitous in almost every field of research. Recently, new research areas devoted to the analysis of graphs and data associated to their vertices have emerged. Focusing on dynamical processes, we propose a fast, robust and scalable framework for retrieving and analyzing recurring patterns of activity on graphs. Our method relies on a novel type of multilayer graph that encodes the spreading or propagation of events between successive time steps. We demonstrate the versatility of our method by applying it on three different real-world examples. Firstly, we study how rumor spreads on a social network. Secondly, we reveal congestion patterns of pedestrians in a train station. Finally, we show how patterns of audio playlists can be used in a recommender system. In each example, relevant information previously hidden in the data is extracted in a very efficient manner, emphasizing the scalability of our method. With a parallel implementation scaling linearly with the size of the dataset, our framework easily handles millions of nodes on a single commodity server.
\end{abstract}


\IEEEpeerreviewmaketitle

\section{Introduction}
The study and application of graph theory have been increasingly active in both the academic world and in industry. The advent of large-scale datasets has lead to the invention of new tools to handle the scale such as graph databases~\cite{angles2008survey} and graph analytics frameworks~\cite{Low_2012, xin2013graphx}. In the academic world, the emerging field of graph signal processing~\cite{Shuman_2013} strives to develop methods combining graphs and data associated to their vertices. 

\begin{figure}[h]
\begin{center}
\includegraphics[width=1\columnwidth]{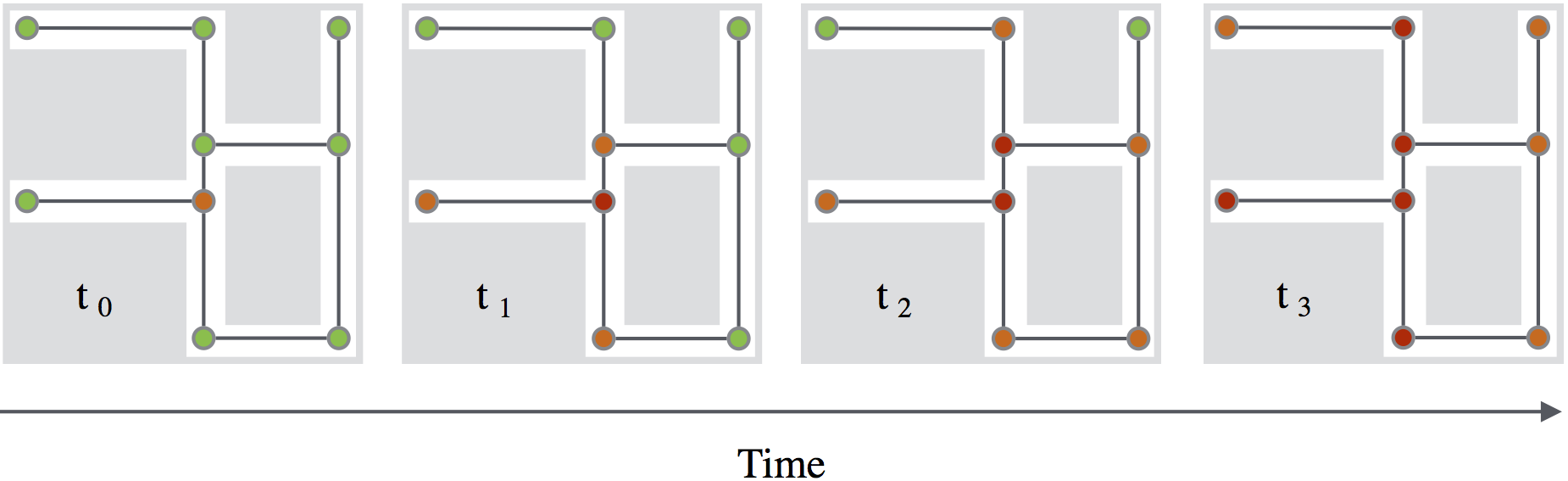}
\caption{{\bf Evolution of a congestion pattern in a localized area of a city represented as a graph.} Each intersection is associated to a node and each road is mapped to an edge. The color of each node represents the concentration of vehicles at this particular intersection (from green, fluid, to red, congested). Different snapshots of the same area show the evolution of the congestion through time. The congestion starts from a single node at time $t_0$ and can only spread over the neighboring intersections through time. Over the days, thousands of patterns can be analyzed to extract important insights such as: how patterns spread, how long they last, are they repetitive, what is the variability of the spread, etc. Learning the particular characteristics of congestion patterns and classifying them would be very beneficial to drivers as appropriate measures could be taken by the authorities as soon as congestion appears.
}\label{Fig1}
\end{center}
\end{figure}

The analysis of dynamical processes taking place on a network is a typical use-case of this combination with applications in various domains such as neuroscience~\cite{beggs2003neuronal, sporns2014contributions}, the study of epidemics in physics~\cite{keeling2005networks,colizza2006role} and rumor spreading in social networks~\cite{de2013anatomy}. In these examples, some quantity or state (such as activity, information or congestion) spreads over a network, as illustrated in Fig.~\ref{Fig1}. Activity patterns formed by these dynamical phenomena or processes are defined by two properties: i) their localization both in space and time and ii) the way they spread, always propagating through the neighborhood over time. In the following, we refer to those particular processes as causal processes. A \emph{causal process} on graph is thus a particular type of dynamical process that models a physical phenomenon propagating and spreading from a node to its neighbors in successive time steps.

In addition, in many applications such patterns appear regularly on the network in the same locations, possibly with some variations. The repetition of the dynamics offers a chance to better understand the underlying process causing the spreading as well as to anticipate or forecasting possible future spreads of a pattern using historical data.

In the present work, we introduce a novel and intuitive method designed to track recurring patterns of activity induced by causal processes on graphs. It relies on the \emph{causal multilayer graph} (CMG), a particular kind of multilayer graph designed to follow the propagation of events in successive time steps (see following section). From the causal multilayer graph and the data attached to its nodes, we extract dynamic activation components, subgraphs of the CMG representing patterns of activity. We then classify and analyze these patterns to reveal global trends and insights on several applications showing that our framework is applicable to a large class of problems. 

{\changes The idea of pattern detection on graphs has some similarity with the temporal motif detection of~\cite{kovanen2011temporal} and the temporal subgraph isomorphism of~\cite{Redmond:2013:TSI:2492517.2492586}. However, motifs are restricted to be subgraphs of a few nodes, typically 2 or 3 (due to the NP-hard graph matching process), while our method can deal with patterns of any size. Our patterns are a type of mesoscale structure within temporal graphs as described in~\cite{holme2015modern}. These structures seem important for understanding dynamical activity in temporal networks but are not well explored yet (except for network communities). We present new results in this direction, as we show in our applications. 
}

{\changes Although there has been a huge amount of research on multilayer networks in the past few years, scalable data-driven methods dedicated to the analysis of dynamic data evolving on multilayer graphs are still lacking~\cite{7093190,holme2015modern}. The proposed method here scales linearly with the number of nodes and time-steps making it possible to handle millions of nodes by leveraging today's multicore architecture. In addition, it is simple to tune as it mainly depends on a single parameter and is very flexible as it supports directed, weighted and dynamic graphs.}

The manuscript is structured as follows. Firstly, we describe the peculiar structure of the causal multilayer graph. Then, we introduce the causal multilayer graph of activity that contains the dynamical patterns to be analyzed. Next, we propose a method to compare and cluster/classify activity patterns which we call dynamic activated components. In the second major part of the manuscript, we illustrate the usage of our framework in three different real-world applications. As a first application we extract dynamic patterns of activity in a social network and compare our approach to the work of De Domenico \textit{et al.} on rumor spreading in~\cite{de2013anatomy}. The second application is devoted to the analysis of the flux of pedestrians and congestion patterns in a train station. Finally, we show interesting ``mood'' patterns extracted from more than $100,000$ collaborative music playlists. {\changes In the Appendix, we dedicate two sections to explain how to construct the CMG of activity efficiently in order to handle large datasets alongside its generalization to dynamic graphs. As part of the open science movement, we release the code and data related to this work to the community under the GPL v2 license on the laboratory's Github account~\cite{kikohs_code}}.

\section{Tracking dynamic activity patterns}
\label{sec:cmg}

{\changes 
In order to track dynamical activity in real data in a computationally efficient way, we need to introduce several mathematical objects and divide the method in several steps. We start by introducing the CMG as a conceptual object to guide the reader and help making connections to previous works on multilayer networks. However in the actual implementation, the construction of the complete CMG is avoided. Instead, we reduce the CMG to small parts by combining the signal describing the activity on the network and the network itself. In this section, we focus on the description of the concept and leave the technical details of the implementation in the appendix~\ref{sec:efficient_construction}. For clarity, we also restrict the presentation to a static graph. However, in the appendix~\ref{sec:generalized}, we present the more general version of our method that takes into account the possible evolution of the graph with respect to time.

Several notations are introduced throughout this section. To help the reader, they are summarized in Table~\ref{tab:notation}.
\begin{table}[h]
\centering
\begin{tabular}{c|c}
 $G$ & Static graph  \\
 $W$, $w_{ij}$ & Weight matrix of $G$ and entry $i,j$ of $W$\\
 $V_G$, $E_G$& Set of vertices and edges of $G$\\
 $K$ & Causal multilayer graph (CMG)\\
 $N$, $T$& Number of nodes per layer and nb of layers of the CMG\\
 $H$ & Causal multilayer graph of activity\\
 $S$, $M$& Signal and mask matrices\\
 $L_t$& Layer $t$ of the CMG\\
 $\Omega$ & Set of intra-layer edges \\
 $\Omega_x$ & Set of inter-layer edges between neighbors on the layer\\
 $\Omega_{xs}$ & Set of inter-layer edges being self-edges\\
 $\mu$ & Threshold value to obtain the binary mask $M$ from $S$
\end{tabular}
\caption{Notations for the different mathematical objects.}
\label{tab:notation}
\end{table}

}
\subsection{The causal multilayer graph}

Multilayer, multislice or multiplex graphs and the applications they model are a topic of great interest with a fairly large literature~\cite{mucha2010community,gomez2013diffusion,Kivela_2014}. A multilayer graph is made of layers (distinct subgraphs) bound together by \emph{inter-layer} edges. Different rules exist for building multilayer graphs. {\changes Our causal multilayer graph belongs to the family of temporal graphs~\cite{Holme_2012}, however with a particular way of connecting layers to account for the causality. To our knowledge, the configuration we propose has only been used in a recent theoretical study on the spreading of an epidemy~\cite{valdano2015analytical}. In that reference, the purpose is purely theoretical. It simplifies the mathematical expressions and allow for new insights on the dynamics of epidemics on temporal networks. In our case, the goal is application oriented: working with this structure leads to a fast method able to track dynamic activities within a (possibly changing) network.}

To construct the CMG two elements are needed. First, a graph that encodes the structural connectivity between the vertices onto which the time-series is defined. Here, we refer to this graph as the spatial graph $G = (V_G, E_G)$. For the sake of simplicity we assume $G$ to be unweighted and undirected although it is possible to apply our method to weighted and directed graphs. $V_G$ is the set of nodes (vertices) with $|V_G| = N$, $E_G = \{(i, j) \,| \, i, j \in V_G\}$ is the set of edges. Second, a matrix $S$ of temporal signals made of $N$ time-series of length $T$, one per vertex of $G$, must be given.

\subsection{Definition}

The causal multilayer graph is made of layers $\{L_0,L_1,L_2,\cdots,L_{T-1}\}$, each one being a distinct copy of the spatial graph $G$ associated to one time-step $t\in[0,\cdots,T-1]$. We denote by $i_t$ the vertex of layer $L_t$ associated to vertex $i$ on $G$. Since each layer is a copy of $G$, the set of \emph{intra-layer} connections in the causal multilayer graph connecting vertices within each layer is the set $\Omega=\{(i_t, j_t,w_{ij}) \,| \, i, j \in V_G, \, t\in[0,\cdots,T-1]\}$, with $w_{ij}$ being the weight of edge $(i,j)$.

To capture the spreading of an event on the graph on successive time steps, each vertex $i_t$ is also connected to its neighbors on $G$ at time step $t+1$. This set of inter-layer edges is denoted $\Omega_x=\{(i_t, j_{t+1}, 1) \,| \, i, j \in V_G, \, t\in[0,\cdots,T-2]\}$. The weight value for the inter-layer edges may be set to an arbitrary value. To simplify and to have an equal treatment of the spatial and temporal dimensions it is here set to 1. 

Moreover, to follow the activity of the same vertex through time each vertex $i_t$ at $L_t$ is also connected to itself, $i_{t+1}$ at $L_{t+1}$ as it is done in temporal graphs~\cite{Holme_2012}. This set of temporal self-edges is denoted $\Omega_{xs}=\{(i_t, i_{t+1}, 1) \,| \, i \in V_G, \, t\in[0,\cdots,T-2]\}$.

\begin{figure}[h]
\begin{center}
\ifdraft
\includegraphics[width=0.5\columnwidth]{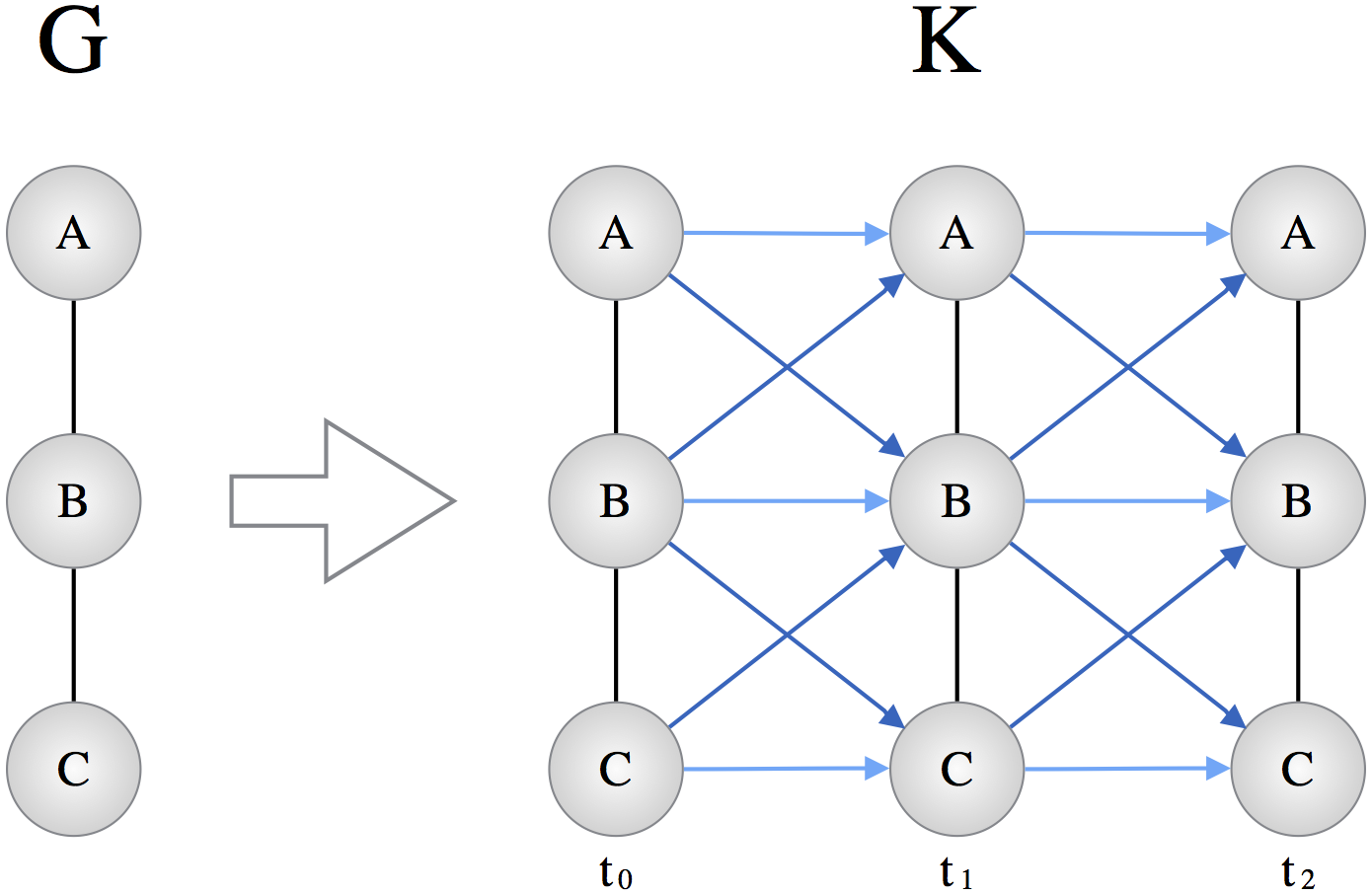}
\else
\includegraphics[width=1\columnwidth]{Fig2.png}
\fi
\caption{
{\bf An illustration of the causal multilayer graph.} The CMG is made of layers: copies of the spatial graph $G$ associated to each time-step ($t_0,t_1,t_2$). The number of time-steps here is $3$. Nodes of the CMG, $K$, are connected by inter-layer edges between two successive time steps if they are neighbors in $G$ (dark blue) or if they represent the same node in $G$ (light blue).
}\label{Fig2}
\end{center}
\end{figure}

In total, the causal multilayer graph $K = (V_K, E_K)$ is composed of $V_K = N\times T$ vertices and of the union of intra-layer, inter-layer and self-edges:
\begin{align}\label{eq:CMGsetfull}
E_K= \Omega\cup \Omega_x\cup \Omega_{xs}.
\end{align}
Remark that in the presented applications, intra-layer edges (the set $\Omega$) are dropped to better capture the propagation of events on the graph and to make clearer visualizations. In the following, $E_K$ is the set:
\begin{align}\label{eq:CMGset}
E_K= \Omega_x\cup \Omega_{xs}.
\end{align}
{\changes \emph{Relation to previous work and alternative definitions.} The CMG $K$ can be viewed as a particular case of a multilayer network defined in \cite{de2013mathematical} via tensor products. Adopting the notations of this reference we can write the adjacency tensor $K_{\beta \tilde{\delta}}^{\alpha\tilde{\gamma}}$ of $K$ as:
\begin{equation}\label{eq:tensordef}
K_{\beta \tilde{\delta}}^{\alpha\tilde{\gamma}}=\sum_{\tilde{h},\tilde{k}=1}^T \sum_{i,j=1}^N w_{ij}(\tilde{h}\tilde{k}) \mathcal{E}_{\beta\tilde{\delta}}^{\alpha\tilde{\gamma}}(ij\tilde{h}\tilde{k}),
\end{equation}
where $\mathcal{E}_{\beta\tilde{\delta}}^{\alpha\tilde{\gamma}}(ij\tilde{h}\tilde{k})$ is the fouth-order tensor of the canonical basis of $\mathbb{R}^{N\times N\times T\times T}$ and $w_{ij}(\tilde{h}\tilde{k})$ denotes the weight of the edge between node $i$ on layer $\tilde{h}$ and node $j$ on layer $\tilde{k}$. In the case of the definition given in Eq.~\eqref{eq:CMGset}, $w_{ij}(\tilde{h}\tilde{k})=0$ when $\tilde{k}\neq \tilde{h}+1$ for all $i,j,\tilde{h}$. Moreover, $w_{ij}(\tilde{h}(\tilde{h}+1))=w_{ij}$ (edges associated to $\Omega_x$) for $i\neq j$ and $w_{ii}(\tilde{h}(\tilde{h}+1))=1$ (edges associated to $\Omega_{xs}$). If like in~\eqref{eq:CMGsetfull} the connections within layers are taken into account, then in addition $w_{ij}(\tilde{h}\tilde{h})=w_{ij}$.

We can also matricize the tensor to have a matrix form representation. In that case the adjacency matrix $W_K$ of $K$ for the definition given in Eq.~\eqref{eq:CMGsetfull} is the following  tensor product of matrices:
\begin{align}
W_K&= I_{T}\otimes W+ O^{(1)}\otimes W +O^{(1)}\otimes I_N,
\end{align}
where $I_T$ is the identity matrix of size $T\times T$ and $O_1$ is the off-diagonal matrix where only the upper first off-diagonal part is non zero, i.e. $O^{(1)}_{i,j}=1$ for $j=i+1$ and zero otherwise. $W$ is the weight matrix of $G$. Again, the first term is associated to $\Omega$, the second to $\Omega_x$ and the third to $\Omega_{xs}$. For the second definition (Eq.\eqref{eq:CMGset}):
\begin{align}
W_K&= O^{(1)}\otimes W +O^{(1)}\otimes I_N.
\end{align}

}
Note that $K$ can also be seen as the strong cartesian product of the graph $G$ with the directed path graph of $T$ vertices, as defined in \cite{sabidussi1959graph}. An illustration of the CMG is given in Fig~\ref{Fig2}.

\subsection{Causal multilayer graph of activity}
\label{sec:act_cmg}

\subsubsection{From signal to binary states} 

In the general case a signal is defined as a set of real values without any priors. However in many applications involving causal processes, the activity over the network is binary: active/not active, infected/healthy, congested/not congested, etc. In the present study, we assume that an arbitrary signal associated to each node of $G$ can be cast in a binary activation vector describing if the node is active or not at a particular time step (i.e by thresholding the signal). We then label each vertex of the CMG with the binary value associated to spatial node $i$ at layer $t$. For example, in the previous traffic illustration (see Fig~\ref{Fig1}) each node of the CMG could be labeled as ``congested/not congested'' by setting a limit value for the number of vehicles at each intersection over which the node is considered congested. 

The way that the matrix $S$ of signals is cast into a binary ``activation'' mask, $\mathbf{M}$, depends on the dataset. It is determined by the definition of the events one wants to track and may involve several application-dependent parameters. In our applications, we use a simple threshold $\mu$, applied after a Z-score normalization of the signal. Let us denote by $S(i,t)$ the value of $S$ at vertex $i_t$ of $L_t$. The entries of $\mathbf{M}$ are given by thresholding the input signal $S$ with a fixed threshold $\mu$:

$$
M(i, t) = \left\{
            \begin{array}{ll}
                    1 \quad\text{if } S(i,t) > \mu, \\
                    0 \quad \text{otherwise.}
            \end{array}
          \right.
$$

\subsubsection{Combining the causal multilayer graph and the binary mask} 
\begin{figure*}%[h]
\begin{center}
\ifdraft
\includegraphics[width=0.75\columnwidth]{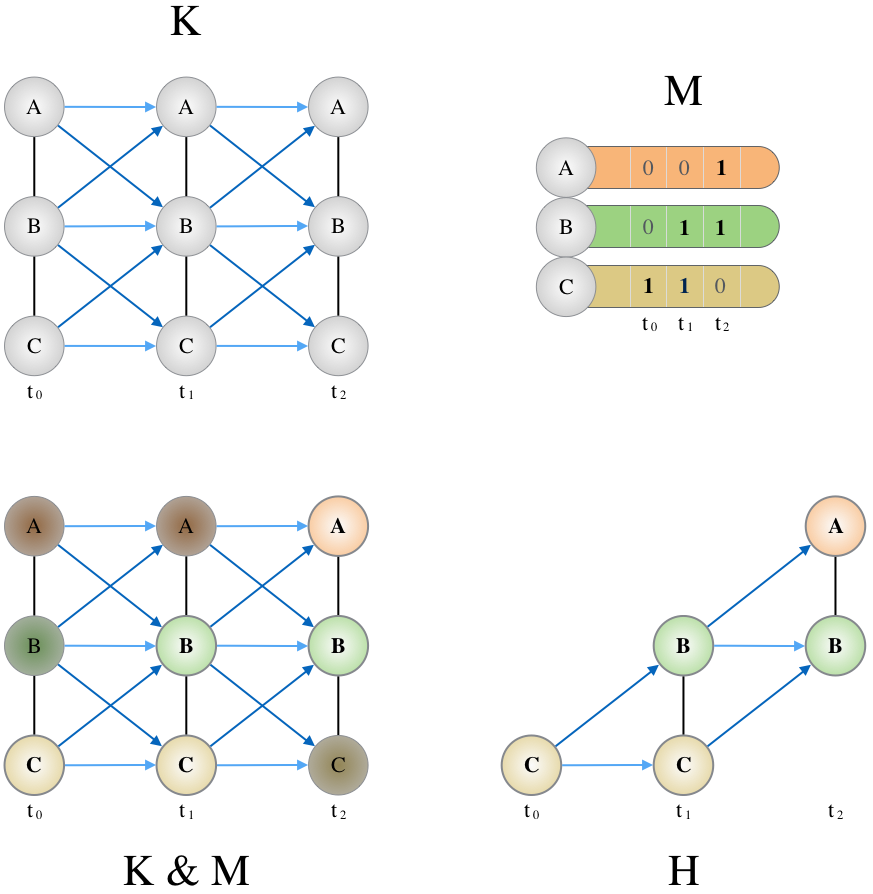}
\else
\includegraphics[width=1\columnwidth]{Fig4.png}
\fi
\caption{
{\bf Construction of the causal multilayer graph of activity $H$.} From the CMG, $K$ (top left), and the mask, $M$ (top right), the CMG of activity, $H$, is constructed by labeling each node of $K$ with the binary state from $M$ (bottom left). This operation is represented by the ampersand `\&' symbol. Then, only the activated nodes are kept in $H$, the rest are discarded (bottom right).
}\label{Fig4}
\end{center}
\end{figure*}

The CMG, $K$, is the skeleton onto which the activation mask is incorporated. From $K$ we create a subgraph, $H = (V_H, E_H)$, called the causal multilayer graph of activity, by taking into account only the set of activated vertices of $K$. We say that $i_t\in V_K$ is activated and hence belongs to $V_H$ if $M(i,t)=1$. Edges of $H$, \emph{causal edges}, exist between nodes if they exist in $K$. In other words, a causal edge exists if both nodes are neighbors on $K$ and activated. To ease the comprehension, the Fig.~\ref{Fig4} summarizes the creation of the causal multilayer graph of activity $H$ from the CMG, $K$, and the binary matrix $M$.

In its current form, the construction of $H$ scales poorly with the size of the data as we need to create $K$ with a number of nodes, $V_K = N\times T$, which can be very large. To optimize this data structure and avoid its full construction, it is more efficient to create $H$ directly from the data. {\changes In Appendix~\ref{sec:efficient_construction}, we describe an efficient, parallelized algorithm, to create the causal multilayer graph of activity. The complexity is linear in the number of edges and vertices of $G$ and in the number of time-steps.}

%%%%%%%%%%%%%%%%%%%%%%%%%%%%%%%%%%%%%%%%%%%%%%%
\section{Analyzing dynamic activity patterns}
\subsection{Dynamic activated components}

We define as dynamic activated components (DACs) the weakly connected components of the causal multilayer graph of activity $H$. Each component, extracted using the standard HCC algorithm \cite{Kang_2009}, encodes an individual pattern of activity induced by the causal process on the spatial graph $G$. For each component, we name \emph{width} the number of layers over which the component spans and \emph{spatial spread} the number of distinct nodes of $G$ contained in the component. Note that subgraphs containing only one node (width equals $1$ and spatial spread equals $1$) are discarded as they add little information to characterize the dynamic nature of an event happening in the data. 

With the exception of grid-like structures, networks do not generally have a regular topology. As a consequence, detecting groups of vertices appearing in a repeated manner for a large number of components proves to be challenging. A first solution could be to compare two DACs using approximate subgraph isomorphism and to cluster them according to their similarity score. Here, we propose a more scalable method that encodes each subgraph as feature vectors. We then rely on a standard clustering algorithm to extract global patterns of activity from DACs (see next section).

The first vector, named static feature vector, is a layer-invariant vector constructed by compressing each layer of the DAC and by counting how many times each node is activated. For example if a spatial node at index $i$ is activated on $3$ layers in one DAC, its value on the vector at index $i$ is $3$. If necessary, the static-feature vector may be normalized using the $\ell^2$-norm to help cluster together components with similar activated nodes but of different temporal width. The idea is similar to the bag-of-word feature vector in text mining. While loosing the dynamical aspect, an activation signature still remains in the static-feature vector. This is enough to perform a meaningful clustering in the applications presented here.

\begin{figure}[h]
\begin{center}
\ifdraft
    \includegraphics[width=0.6\columnwidth]{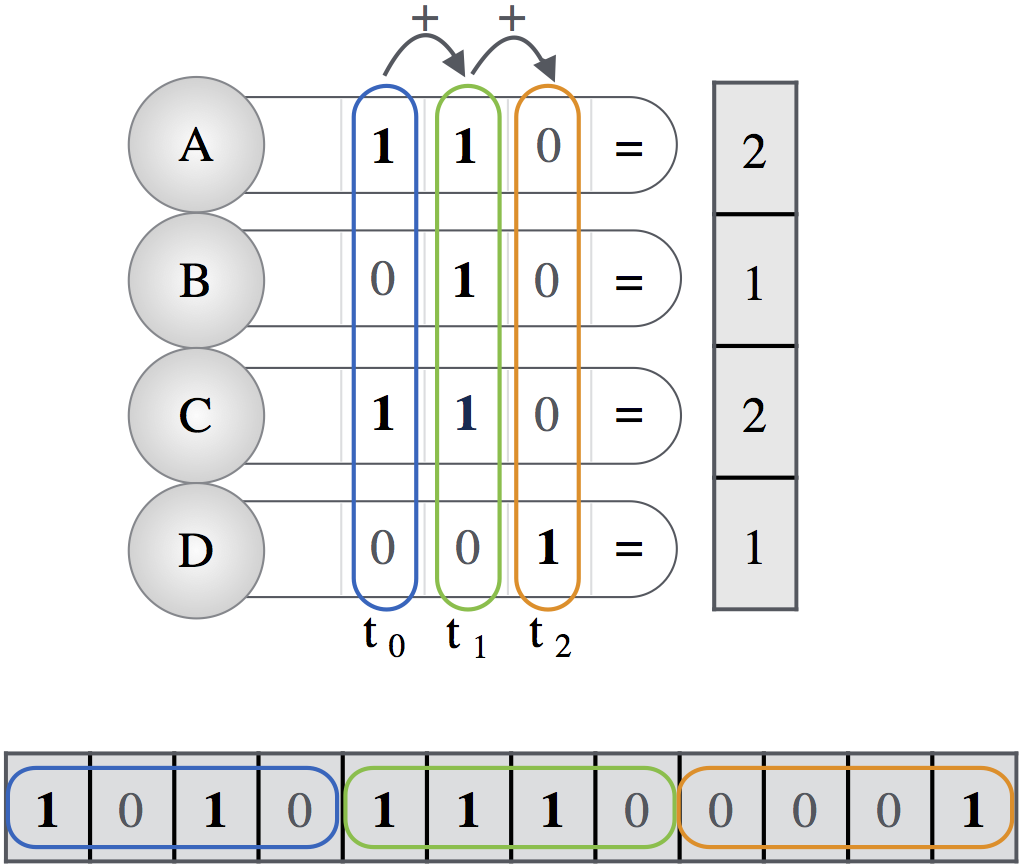}
\else
    \includegraphics[width=1\columnwidth]{Fig7.png}
\fi
\caption{
{\bf Creation of static and dynamic feature vectors.} On the right side, the static feature vector is obtained by compressing the layers and counting the occurrence of each node. On the bottom, the dynamic feature vector is created by vectorization of the component's activation mask.
}\label{Fig7}
\end{center}
\end{figure}

In addition to the static features, we also encode each DAC as a dynamic feature vector. In essence, it is just a vectorization of the activation mask of each active component. It contains all the information concerning the dynamic arrangement of the component while being memory friendly and computationally efficient. These vectors will be used later on to extract a common representative pattern from each cluster (see analysis of cluster properties). To get a clearer picture of these two feature vectors we refer the reader to Fig.~\ref{Fig7}.

\subsection{Clustering the components}

Clustering DACs is a crucial step in our method as it reveals meaningful groups of components which share common characteristics. In most cases, the groups are unknown and the clustering is thus unsupervised. While our method is not tied to a specific algorithm, the large number of DACs imposes a fast clustering method. We choose the $k$-means algorithm as it scales nicely with the load in quasi-linear time using Llyod's method \cite{Lloyd_1982}. This algorithm partitions the dataset into $k$ sets, $\{A_1,\,A_2,\,\ldots,\,A_k\}$ so as to minimize the within-cluster sum of squares:
\begin{equation}
\sum_{\ell=1}^{k} \sum_{ \alpha_i \in A_{\ell}} \left\| \alpha_i - \overline{\alpha}_{\ell} \right\|^2_2,
\end{equation}
where $\overline{\alpha}_{\ell}$ is the centroid of $A_{\ell}$. We expect here $k$ distinct types of DACs that are repeating themselves, with copies possibly differing by a few nodes (due to noise or other activation behaviors). Using the Lloyd's method, the complexity is proportional to $N_cNki$ where $N_c$ is the number of DACs and $i$ is the number of iterations before convergence. In practice this number is small, so that $k$-means can be considered to scale linearly.

\subsubsection*{Choosing the right $k$}

{\changes 
Automatically choosing $k$ has fueled decades of research in cluster analysis~\cite{arbelaitz2013extensive}. While a universal solution to alleviate this issue remains to be found, several methods propose to estimate the number of clusters~\cite{Bischof1999, pelleg2000x, sugar2003finding, frey2007clustering}.
Particularly, the silhouette width~\cite{rousseeuw1987silhouettes} is a data-driven method that can give a good estimation of the number of clusters in various datasets~\cite{arbelaitz2013extensive, lleti2004selecting}. However, the automatic detection of a unique $k$ has some limitations. Indeed, a community structure can exist at different scales within the same dataset. Since the silhouette coefficient provides only one $k$, it sets the scale of the observation to a unique level of details.
In this work, the choice of $k$ is driven by physical considerations on the data, see the applications on traffic and music data in Sec.\ref{sec:applications}. This allows us to work at a desired scale that is physically meaningful and provides interpretable results. More precisely, we first estimate the order of magnitude of $k$ (how many clusters should we expect? 10, 100, 1000?). We then compute several clusterings for different $k$ at this chosen magnitude and study the resulting clusters for validation \emph{a posteriori} (see Sec.\ref{sec:applications}) .In addition, we check the robustness of the clustering: if a small variation of $k$ involves a large variation in the clusters obtained, the clustering is not reliable and another method must be used.}

\subsection{Analysis of the cluster properties and average activation component}

By grouping similar DACs, the clustering exhibits structure and proves to be invaluable in the analysis of the underlying causal process. However, since we use static-feature vectors as input of the clustering algorithm, the dynamic activity of each DAC is lost in the ``compression'' and cannot be retrieved using the $k$-means centroids. To go beyond the mere analysis of the centroids, we propose to create average activation components (AAC) to represent the average dynamic activity of each representative pattern. To do so, we use the dynamic feature vectors associated to the DACs of a given cluster. For each layer we compute the number of times each node appears for all these DACs. We proceed similarly for causal edges. Note that the choice of the clustering technique used to assign each DAC to a cluster number is irrelevant here as we only need the cluster number to create an average activation component. 

From nodes and causal edge counts, we compute node and edge likelihoods in each cluster and store them as node and edge weights respectively. For a node, it represents the likelihood of activation at a particular layer. For an edge, it represents the chance of a node on the current layer to get activated as a consequence of a node activated on the preceding layer. The final step consists in sparsifying the AAC by discarding nodes and edges with low weights (typically $<5\%$). It removes a significant part of the noise induced by the clustering of the underlying components. For a typical usage of AACs, see the third application ``Analyzing thousands of collaborative audio playlists''.

\section{Applications}
\label{sec:applications}

In the following sections, we test our framework's ability at retrieving and analyzing recurrent patterns of activity, induced by causal processes on a graph, in real applications. We also illustrate the versatility of our framework by creating causal multilayer graphs of activity and extracting dynamic activation components on three completely unrelated datasets, showing that it can be applied to a large class of problems. We nevertheless delay the complete study of each example in future work as it would be beyond the scope of this paper. For clarity purposes the causal multilayer graph of activity is now simply named causal multilayer graph.

\subsection{Rumor spreading on twitter and dynamic activity of communities}
\label{sec:higgs}

Our first application aims at revealing the dynamic activity of communities when a rumor spreads in a social network. The dataset, available at SNAP~\cite{snapnets}, contains the Twitter activity around specific hashtags related to the Higgs boson discovery by CERN in 2014. This relatively large dataset, with a graph of more than ${0.4}$ million nodes and ${14}$ million edges, has been studied in~\cite{de2013anatomy} in order to understand rumor spreading in networks. The size of the dataset is considered as a test for our parallelized algorithm and its ability to scale, as it only takes a few seconds to extract DACs on a community server with 24 cores. The code used for the analysis of this dataset is available online~\cite{kikohs_codeHiggs}.

The Twitter activity has been recorded before, during and after the announcement of the discovery of the Higgs boson by CERN on the 4th of July 2014 at 8:00 AM GMT. The recording starts from the 1st of July and lasts until the 7th. The authors have recorded the Retweet, Reply and Mention activity containing selected keywords related to the Higgs boson. The graph of Twitter followers is provided together with the activity of the users during the event. To study the dynamical activity, \cite{de2013anatomy} takes advantage of the fact that the activity can be tracked over time using retweets. A retweet, in addition to giving a timestamp, provides the causal link between two users. A user retweets an information as a consequence of a first user having tweeted the information. The cascade of retweets can be followed without relying on a causal multilayer approach. Note that the retweet information is asynchronous (not having regular time steps) and can not be directly cast into a causal multilayer graph. It would involve connections between layers not necessarily adjacent. The analysis of the retweet dynamics allowed the authors of~\cite{de2013anatomy} to reveal the bursty behavior of retweet chains, in particular around the official announcement. 

As a proof of concept, the first goal in this example is to confirm the retweet dynamic made of bursty events discovered in~\cite{de2013anatomy}. If this is so, the DACs should be large especially around the announcement. The second task is to show that the activated components can bring new insights on the dataset and on rumor spreading mechanisms. Since the dataset is related to a single \emph{extraordinary} event we do not expect to find repeated patterns of activity. Nevertheless, the analysis of repeated patterns could be done in future work with a larger recording of twitter activity, containing different events appearing over several weeks.

To build the causal multilayer graph we use the activity over time (time series) and combine it with the graph of followers (the social network) as follows.

\subsubsection{The graph}
The Twitter follower graph is a directed network where a user (source) is connected to another (destination) if he is followed by him/her. The graph is made of $456,626$ nodes of users who have been active (retweet, reply, mention) at least once during the recording and more than $14$ million edges.
\subsubsection{The signal and the mask}
To track the activity over time and users, we cut the Twitter recording into regularly spaced time steps. Within a time step, a user is active if he/she has retweeted\footnote{We do not include replies and mentions to be able to compare with~\cite{de2013anatomy} which focuses more on retweets than the other activities.} about the Higgs boson. Different activity patterns may appear at different scales of time (from seconds to hours to days) and the time step is chosen in order to select a particular scale. 
Remember that connections within the causal multilayer graph are only allowed within layers and between two successive layers. The retweet action of a user may occur after several time steps and this is not taken into account in our construction. However, there is evidence of bursty behaviors in social activity~\cite{barabasi2005origin,karsai2012universal} and in particular in this dataset~\cite{de2013anatomy}. A retweet is likely to be done shortly after a tweet appears in the user feed and the likelihood of retweeting decreases with time (following a power law). Hence the time series configuration captures most of the dynamic activity. Moreover, it is also of high interest to focus on tracking the activity solely due to the bursty behavior. Of course the time step length must be chosen in order to match the time scale of the bursty processes: a length of 1 minute is a reasonable choice according to the results of~\cite{de2013anatomy}. On the fourth panel of Fig.5 in their paper, the curve shows a maximum of retweets having a time delay of 1 minute and the number of retweets drops exponentially when the delay increases. 
{\changes{To highlight the impact of such a choice, we run a second analysis with a 10 minute step}}. On the curve the number of retweets within a 10-minutes delay is already 2 order of magnitude smaller than the 1-minute delay.

\subsubsection{Analysis of the activated components, recovering the results of the literature}

The activated components are directed layered graphs of activity. Each node represents a user, active at a particular time-step. Only inter-layer edges have been kept for the construction of the components. On table~\ref{tab:10minutes}, the largest activated components extracted from the causal multilayer graph are shown (10 min sampling). The largest one appears on the 4th of July (the announcement day) and covers most of the day. Moreover, the number of users involved is extremely large. It shows how the information has spread over the network. It starts before 8:00 a.m. as rumors and discussions on the topic spread before and increase as the announcement time approaches. The other components are at least one order of magnitude smaller and last one to three hours each. They are distributed between the 2nd and 5th days of July. Bursts of activity may appear indistinctly during the day or night as the event is of worldwide importance. The geo-localization tags are not available in the dataset and we could not verify whether components involve particular countries or regions.

\begin{table}[h]
\centering
\caption{Largest activated components with their size and time of appearance for the 10 minutes time-steps.}
\label{tab:10minutes}
\begin{tabular}{c|c|c|c|c|c}
	\# Nodes &	\# Layers	& Social spread & Start &	End\\\hline
	55037&	108	&36800&	04, 03:10&	04, 21:00\\
357	&15	&324&	03, 17:40&	03, 20:00\\
	299&	12	&277&05, 11:30&	05, 13:20\\
	254	&12	&231	&05, 05:00	&05, 06:50\\
	244	&9	&235&	05, 00:00	&05, 01:20\\
	232&	12&	212&02, 16:20&	02, 18:10\\
200&	14	&163&	03, 21:00&03, 23:10\\
	169	&5	&107&	04, 14:40&	04, 15:20\\
	166	&9	&160&05, 10:30&05, 11:50\\
	142	&9	&128&04, 20:50&04, 22:10
\end{tabular}
\end{table}

The largest activated components obtained by the one minute sampling are displayed on Table~\ref{tab:1minute}. The largest components appear on the 4th, as expected. In the first component, $4704$ different users are active and the activity spans over $114$ minutes. This frenetic activity appears \emph{just before} the official announcement. The second largest component is the consequence of the announcement, it starts at 8:01 a.m. and the information propagates quickly until 8:18 a.m. Most of the largest components last around 10-15 minutes and involve around a hundred users. The largest components take place on the 4th, where the activity is so frenetic that information can be retweeted in less than one minute and propagated to tens of users in only 10 minutes.

\begin{table}[h]
\centering
\caption{Largest activated components with their size and time of appearance for the 1 minute time steps.}\label{tab:1minute}
\begin{tabular}{c|c|c|c|c|c}
	\# Nodes &	\# Layers	& Social spread & Start &	End\\\hline
8593&	114	&4704&	04, 05:17	&04, 07:10\\
	255	&18	&214&04, 08:01	&04, 08:18\\
216&	15&	164	&04, 04:51	&04, 05:05\\
151&	9	&130	&04, 05:09	&04, 05:17\\
	142&	5	&133&	04, 13:41&04, 13:45\\
	100&	9&	83&	04, 07:13&	04, 07:21\\
	98&	15&	98&	04, 13:22&	04, 13:36\\
	95&	15&	75&	04, 15:08&	04, 15:22\\
	95&	14&	95&	02, 19:57	&02, 20:10\\
	93&	16&	91&04,	14:12&	04, 14:27
\end{tabular}
\end{table}

\subsubsection{Analysis of the activated components, new results, evidence of dynamic communities of active users}
The retweet activity does not fully account for the spreading of rumors. {\changes For example a user may see several tweets concerning the Higgs boson in their feed and decide to retweet only one of them, or possibly decide to come back to the initial source of information to retweet it. In another case, he can tweet the information without mentioning any source.} In these cases, the action of tweeting does not take into consideration the full user network even if it has a clear influence on him/her. Actually, we have compared the graph of followers to the graph obtained by connecting users according to the retweet data. Only 59\% of the edges of the retweet graph match the followers graph: a large portion of the retweets are not from direct neighbors. However, by accounting for the neighbors influence on users, our causal multilayer graph approach reveals the existence of communities of users appearing dynamically as the information spreads over the network. We provide evidence for this claim in the following.

First, we remark that the number of layers (time spread or width) of the two different sampling rates are similar: the largest component in each is around $100$ layers long while the others are around $15$. The number of layers is the number of time-steps, so components of the 10 min sampling rate last $10$ times longer than the 1 min sampling rate. For these two cases there seem to be a scale invariance in time. However, the social spread (number of different users in the component) is less than $10$ times larger between Table~\ref{tab:10minutes} and Table~\ref{tab:1minute}. With the exception of the largest component, it is only multiplied by $2$ (roughly). These two facts tend to advocate for a community-like activity where information is retweeted within communities, limiting the number of users involved.

\begin{figure}[h]
\begin{center}
\includegraphics[width=1.0\columnwidth]{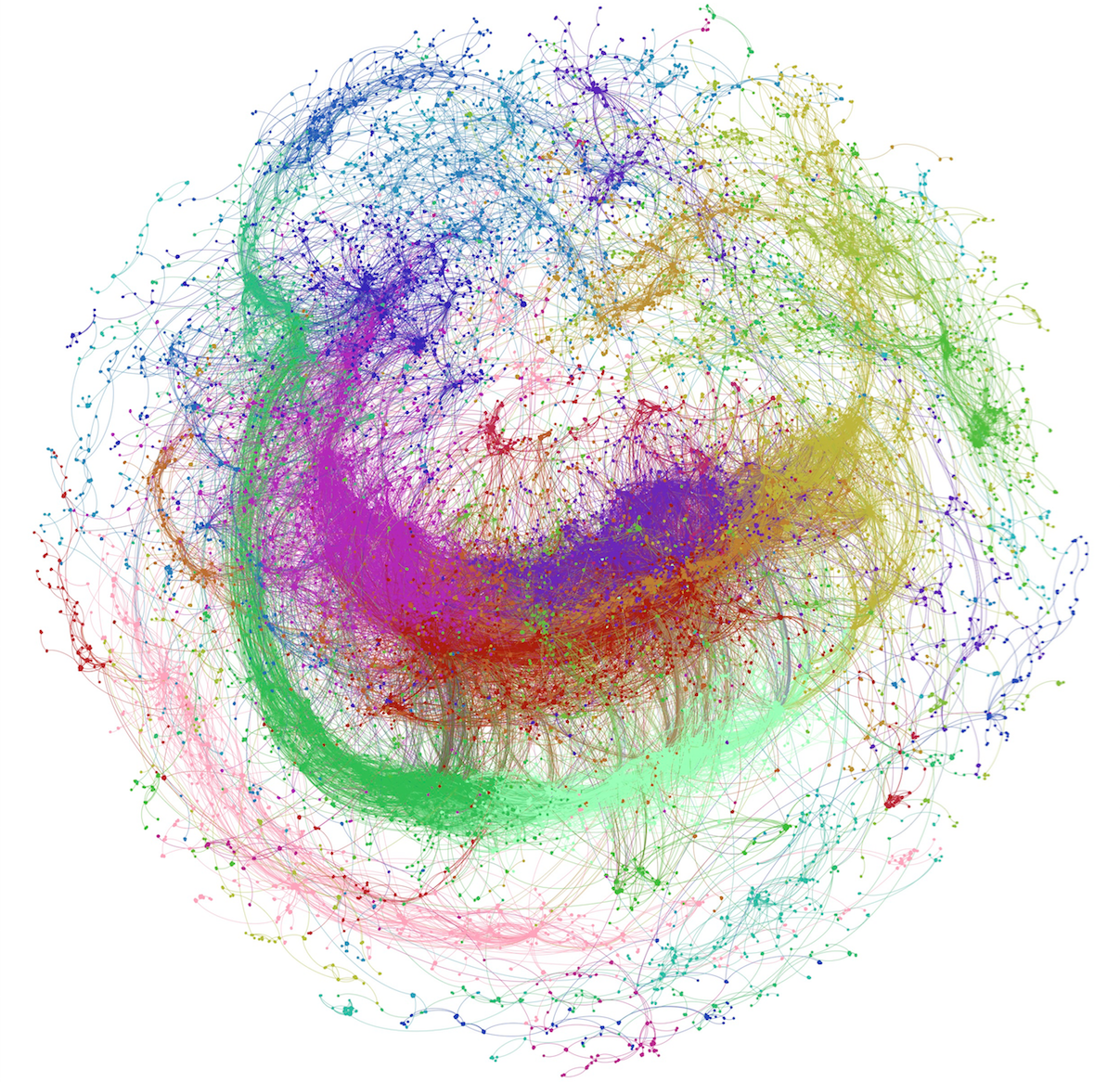}
\caption{ {\bf Graph of the largest activated component from the 10 minute sampling, colored by community.} The graph layout has been generated using open ORD~\cite{martin2011openord}. The different colors correspond to the different communities. However, the set of colors is limited and different communities may have the same color. Edge colors correspond to the color of the source nodes they connect to.
}\label{Fig17}
\end{center}
\end{figure}

In order to better understand the rumor spreading dynamics, it is also interesting to focus on the largest activated component on the 10 min sampling rate, described on the first line of Table~\ref{tab:10minutes}. This component covers the official announcement of the Higgs boson discovery. As each DAC is a graph, it is possible to run a community detection algorithm on it. {\changes Using Louvain's method~\cite{Blondel_2008}, we obtain} $56$ different communities\footnote{Due to anonymization of the data and the non availability of the geo-localization data we were not able to check if these communities correspond to a particular location or group of people.} with $4$ main ones containing a large number of nodes (11\%, 9\%, 7\%, 7\% of the total number of nodes respectively). A plot of the graph with the different well connected communities is shown on Fig.~\ref{Fig17}. 

Communities are elongated as connections in the graph exist between successive layers only: 2 nodes in the same community may not be directly connected but are connected through their neighbors at successive time steps. We recall that each node of the graph is a user active at a particular time-step. Hence several nodes can correspond to the same user and communities in the activated component do not necessarily represent communities in the twitter graph of followers. Communities in DACs are thus sets of connected users \emph{active within a particular time interval}.

\begin{figure*}[t!]
\begin{center}
\ifdraft
\begin{tabular}{cc}
    \includegraphics[width=0.45\columnwidth]{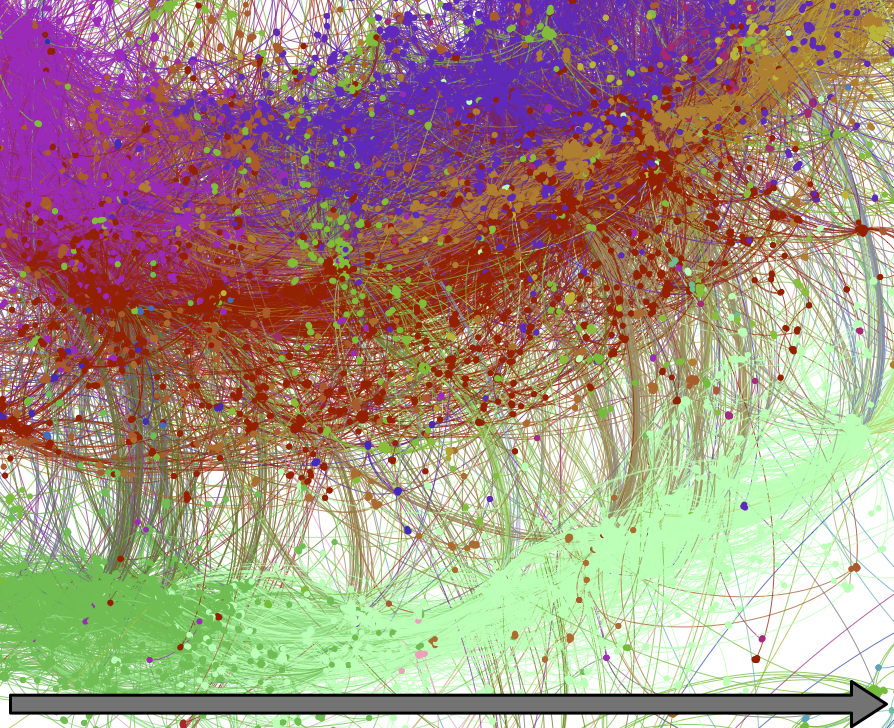} &
    \includegraphics[width=0.45\columnwidth]{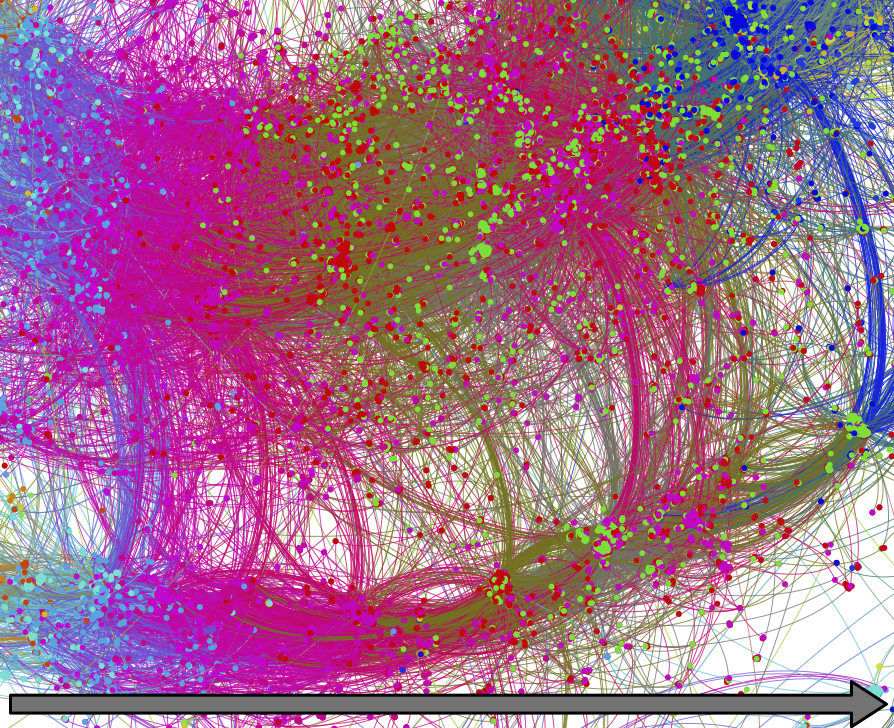}
\end{tabular}
\else
\begin{tabular}{cc}
    \includegraphics[width=0.7\columnwidth]{Fig8.png} &
    \includegraphics[width=0.7\columnwidth]{Fig9.png}
\end{tabular}
\fi
\caption{{\changes{\bf Zoom in a region of interest from the Higgs CMG of activity}. The arrow represents the evolution of time. Edge colors correspond to the color of the source nodes they connect to. The two images represent the same part of the network of Fig.~\ref{Fig17} but colored differently. On the left, the graph is colored by communities (as for Fig.~\ref{Fig17}). On the right colors represent different and successive time-steps of 10 min. The time increases following the color order: light blue, pink, brown, dark pink, dark blue (from left to right). These images reveal a part of the rich inter-community dynamics. On the left image, elongated communities evolve over time. We can see interactions between communities: some are tightly connected (dark pink, purple and red), some are weakly connected but active in the same time (green and dark pink or light green and purple). There exist bursts of connectivity between communities appearing over time (between the green community and the others for instance).
}} \label{Fig:higgs}
\end{center}
\end{figure*}

{\changes A part of the dynamical activity of the component is plotted on Fig.~\ref{Fig:higgs}. On the left, the graph is colored by communities similarly to Fig.~\ref{Fig17} whereas on the right, each color corresponds to a time-step (a layer) instead of a community. By comparing the two graphs, we notice that communities span across layers. Different communities can be active at the same time (purples and greens on the left) while some of them activate in a successive manner (dark green to light green on the left) leading to \emph{spatio-temporal communities} of users. Some communities are tightly connected on several successive layers (red and purple on the left figure) while some others seem to be interacting by bursts of connections over time (red and greens).
This phenomenon seems very interesting and definitely deserves further investigations. New visualizations are needed to depict more precisely these phenomena.
}

\subsection{Visualizing crowd movements in a train station}

Our second application visualizes and quantifies pedestrian movements in the train station of Lausanne, Switzerland. The dataset, gathered by Alahi in~\cite{Alahi_2014}, is composed of 42 millions points (x, y, t, pedestrian id) tracked by a connected network of cameras placed in the two main corridors of the station. The data collection spans over 2 weeks at the most crowded hours {\changes (7-8 am and 5-6 pm)}.
Our study aims at characterizing repeated congestion patterns over time while understanding their dynamics. Far from a purely academic interest, the Swiss Federal Railways are seeking to expand the corridors in prevision of an ever increasing traffic. This work constitutes a first step in the visualization of repeated crowd movements in the station.

\subsubsection{The graph} 
\begin{figure}[h!]
\begin{center}
\includegraphics[width=1.0\columnwidth]{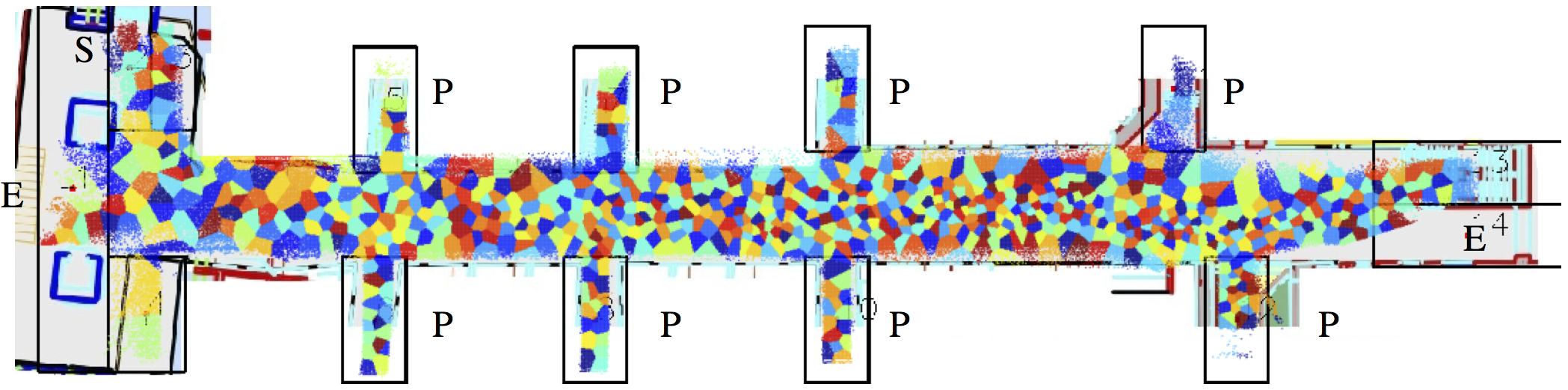}
\caption{ {\bf Segmentation of the west corridor into regions of interests according to the density of pedestrians.} On the corridor map, the 8 black boxes labeled as `P' represent the platforms to access the trains. The letter `E' stands for the two exits and `S' is a shop. Each colored polygon is a small area of the corridor, represented by a single node on the graph.
The main access to the station, on the right, is busier than the one on the left, giving smaller polygons.%
}\label{Fig10}
\end{center}
\end{figure}

Before the creation of the causal multilayer graph, a spatial graph of connected regions onto which the crowd moves needs to be created. In order to facilitate the interpretation of the results, the station corridors can be divided into small areas of one square meter. The number of persons over time passing on each of these areas directly gives the congestion rate in person per square meter. However, some locations are more important than others as the crowd does not evenly spread over the corridors. To account for the crowd's density, we use an adaptive algorithm that robustly cuts the space into fine-grained areas on the most congested zones and into coarser areas where the traffic is less dense. In addition, areas are constrained to be within the same range of surface, around 1 meter square. {\changes This adaptive grid is more precise where the congestion flow needs to be monitored and controlled. Less dense regions are represented by coarser areas leading to more computationally effective algorithm on those parts of the grid.} The result can be seen on Fig.~\ref{Fig10}.
This technique is commonly used in computer vision applications to cut images into ``Superpixels''. The method we use is detailed in~\cite{Achanta_2012} where the pixel colors are here replaced by the total number of persons who crossed each spatial point. The nodes of the spatial graph $G$ are the superpixels and the edges are created by linking adjacent superpixels.

\subsubsection{The signal and mask} We naturally choose the pedestrian's density in the station as the signal. The signal sampling rate is set to 5 seconds. This order of magnitude for the time sampling emphasizes the slow movement of a large crowd (congestion) over the faster normal flow (5 sec is the average time needed for a person stuck in a congestion to move from one superpixel to an adjacent one). The value of the signal at each time step is the number of pedestrians that have crossed the area within this 5 sec duration. 
{\changes We normalize each signal to have a zero mean and a standard deviation of one (Z-score) as we are more interested in the variations rather than the absolute number of pedestrians. we create the congestion mask $M$ by empirically setting the threshold $\mu$ and applying it to the normalized signal. This casts the traffic activity in a congested/non congested state.} 
The width and spatial spread of DACs are directly related to the threshold value (higher values give smaller components). In this application, the best threshold ($\mu=1$, one standard deviation of the initial signal) is chosen so that the spatial spread of the DACs are on average within the range of the distance of one exit to another. This choice of the threshold is relevant to model pedestrian trajectories as it emphasizes on the crowd direction and global behavior within the station.

\subsubsection{Analysis of the activated components} The causal multilayer graph is made of thousands of components naturally split by the activation threshold. The basic statistics of each component, such as width and spatial spread, are useful to quantify the impact of an event (e.g a train departure) in the whole station. The width gives the duration of an event and the spatial spread the number of regions impacted by it. Note that the congestion event is tracked over time, the width and spatial spread take into account all of the congestion patterns even if it moves along the corridors.

\begin{figure}[h!]
\begin{center}
\includegraphics[width=1.0\columnwidth]{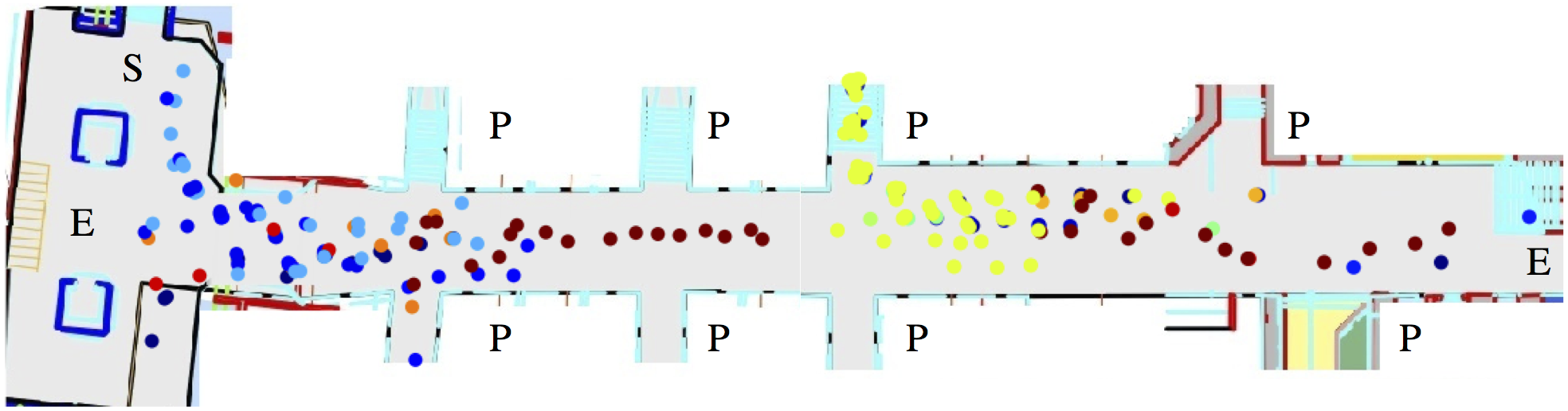}
\caption{ {\bf Two examples of average activated components plotted on the same figure.} Each circle represent a crowded area at some point in time. The color scale gives the time of appearance of a congestion at a given location. It ranges from dark blue (start of the congestion) to yellow, then to red for the end of the component. On the right is an average activated component giving a repeated congestion pattern between one platform and the main exit. At the beginning of the component pedestrians go on the platform (blue dots under the yellow), just before a train departure. Then the crowd coming out of the train can be tracked, from yellow (earlier time) near the platform access to red (later time) as it approaches the main exit. The second activated component on the left shows the crowd coming both from the left entrance and from a shopping store and entering the main corridor. This component ends inside the corridor as the crowd then spreads to different platforms, reducing the congestion rate below the threshold.
}\label{Fig11}
\end{center}

\end{figure}

The k-means clustering of thousands of components of similar shapes creates average activated components correlated with the departures and arrivals of trains in the station. 
{\changes To choose an appropriate $k$ we proceed as follows. Knowing that the time-series span over two hours per day, the expected number of departures and arrivals is between 15 to 25. This gives an order of magnitude for $k$ since we expect clusters to be correlated with arrivals and departures of trains. We then compute the clustering for different $k$ around these values. With $k=20$, clusters have flow patterns with a spatial length of several meters, representing accurately pedestrians trajectories within the station.}
Two examples are given on Fig.~\ref{Fig11} (left and right). Each one represents an average dynamic trajectory of pedestrians inside the corridors. On these average components, a node represents an activated (crowded) area of the corridor, the node color represents the time dimension, from the start of the component (blue) to its end (red). The two examples display mean congestion patterns evolving in time as the crowd moves along the corridor. It demonstrates the ability of our causal multilayer graph model to track the crowd movement and extract relevant information from it.

The average activated components (associated to each cluster) can be used as a basis for the analysis of the most recurrent events in the station. It provides information such as the most frequently crowded areas or the largest congestion in time, on space, highlighting any traffic ``bottleneck''. In addition, unusual events (delay, accident, etc.) in the station causing a congestion can be detected by comparing its activated component to the average one of each cluster. A large dissimilarity with all the clusters is considered abnormal and may require a human intervention in the station.

%%%%% AUDIO

\subsection{Analyzing thousands of collaborative audio playlists}

Up till now, the presented applications have extracted and analyzed activated components from causal processes modeled as time-series on a graph. In this application, we use the causal multilayer graph approach to analyze another kind of dataset without time series but with causal relations between nodes of the spatial graph. Our objective is to show how to use activated components to create a playlist recommender system based on what Bonnin \textit{et al.} describe as frequent pattern mining in~\cite{Bonnin_2014}. For illustration purposes we also visualize groups of common listening patterns of users. This example shows that a wide range of applications can be modeled by our method. Note that the focus here is on the method (how to construct a model from causal data, how it scales) and not on the results, there are probably better ways to build recommender systems for music.

The Art of the Mix dataset originally crawled by McFee in~\cite{mcfee2012hypergraph} regroups $101,343$ collaborative mixes from 1998 to 2011. A mix is a special kind of playlist where songs are chosen to have meaningful transitions between them. In other words, songs are put in a specific order in a mix because there exists a causal relationship between them. The position inside the mix is also important as a mix possesses a global evolution. The types of songs (their mood, energy, degree of danceability) in the first part are often different from the ones in the middle or at the end. In addition to the ordering of songs, a mix is also associated to a playlist category by a user. These playlist categories such as: ``Rock'', ``Romantic'', ``Single Artist'', etc, help to navigate between the thousands of playlists on the site. This information is used in~\cite{mcfee2012hypergraph} to validate their approach.

\begin{figure*}[t!]
\begin{center}

\ifdraft
    \begin{tabular}{cc}
    \subfloat[Metal component. While the genre diversity is a bit higher than in the Jazz example (b), all genres belong to the same meta-genre: Metal.\label{Fig12}]{\includegraphics[width=0.45\columnwidth]{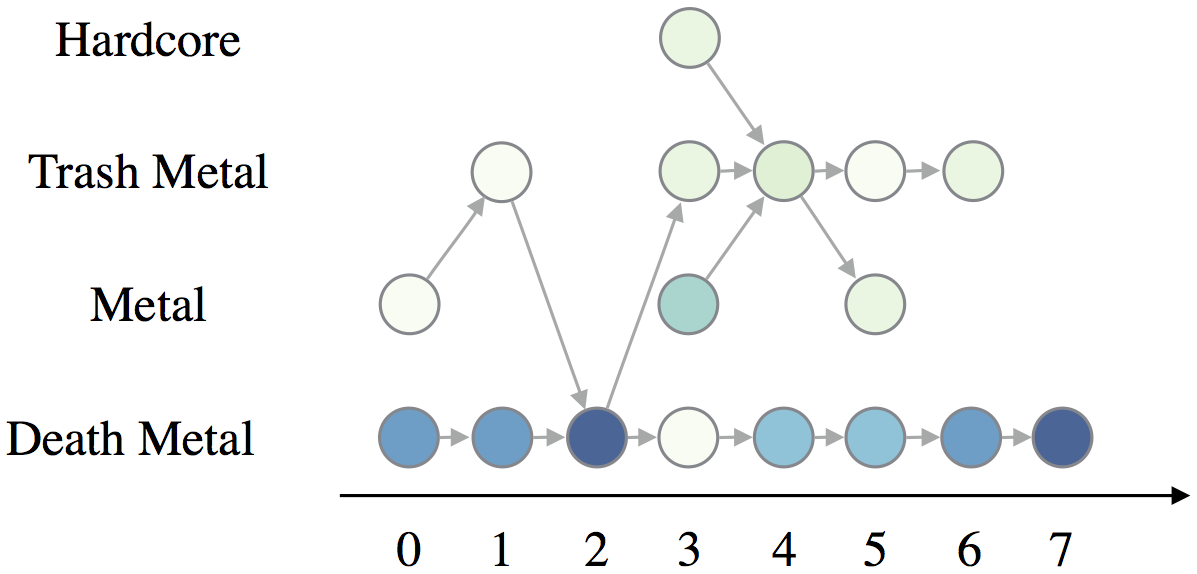}} &
    \subfloat[Jazz component. We clearly see that Jazz playlists are very pure and do not mix with other genres. The same phenomenon also exists for other music niches such as Classical music.\label{Fig13}]{\includegraphics[width=0.45\columnwidth]{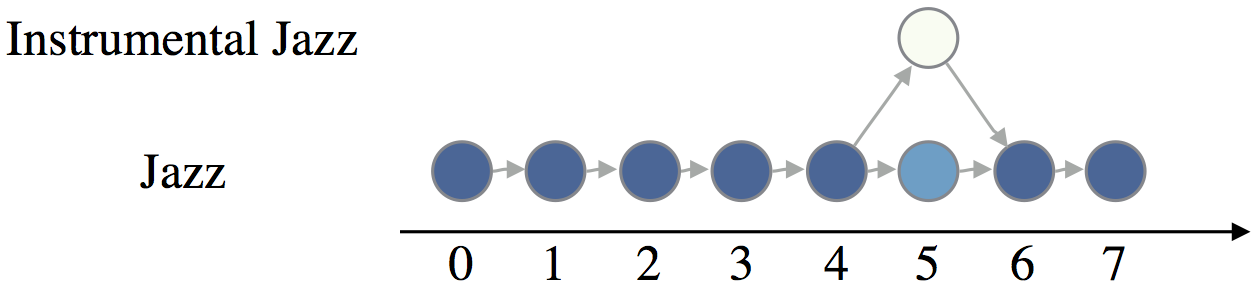}} \\
    \subfloat[Rock - Alternative component. The diversity of the genres is much higher than in previous examples but stays coherent.\label{Fig15}]{\includegraphics[width=0.45\columnwidth]{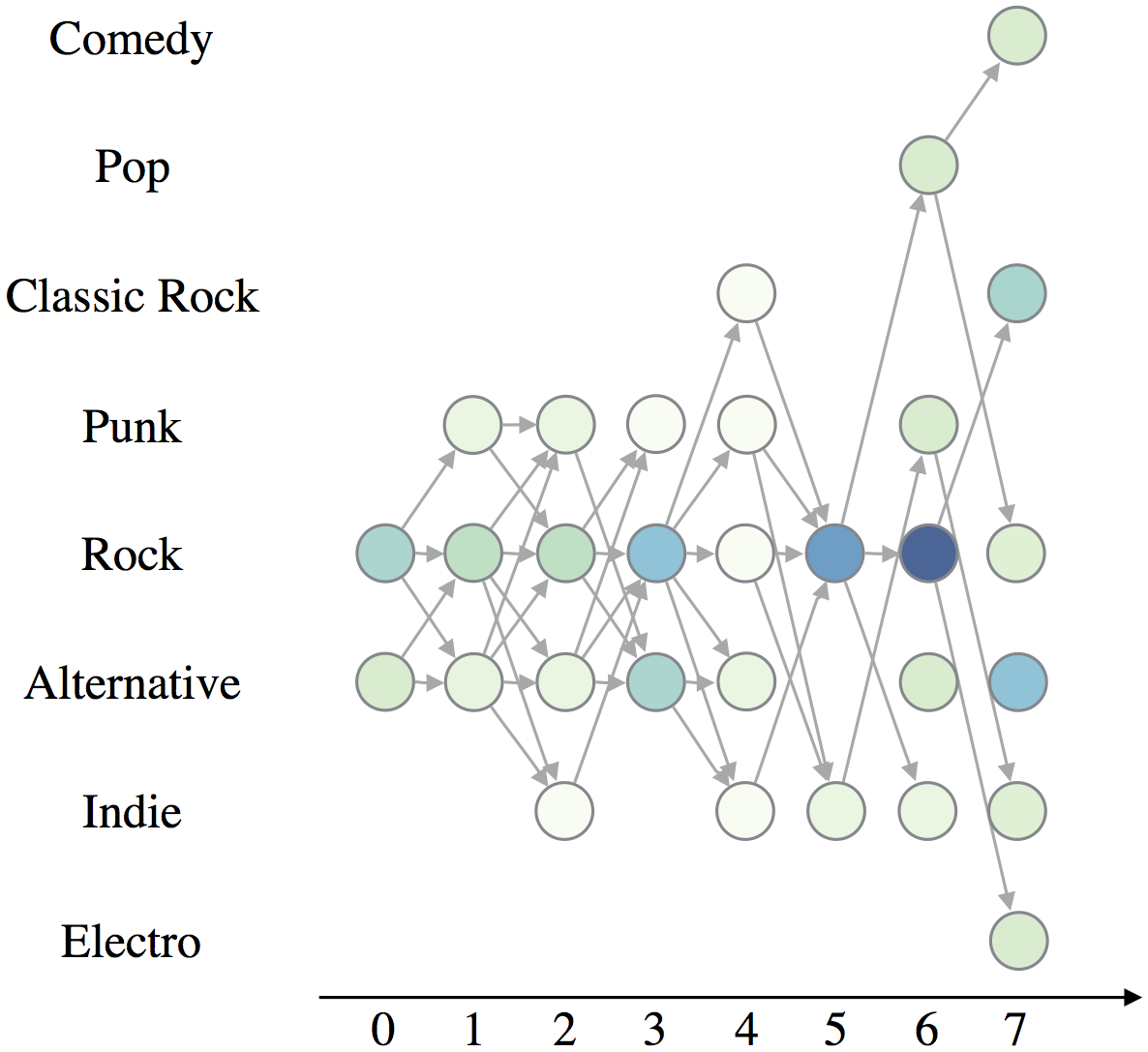}} &
    \subfloat[Rap / Hip-hop component. Heavy Metal nodes are not outliers but are indeed connected to Hip-Hop with songs from famous artists such as Korn, Limp Bizkit, Public Enemy, etc. \label{Fig14}]{\includegraphics[width=0.45\columnwidth]{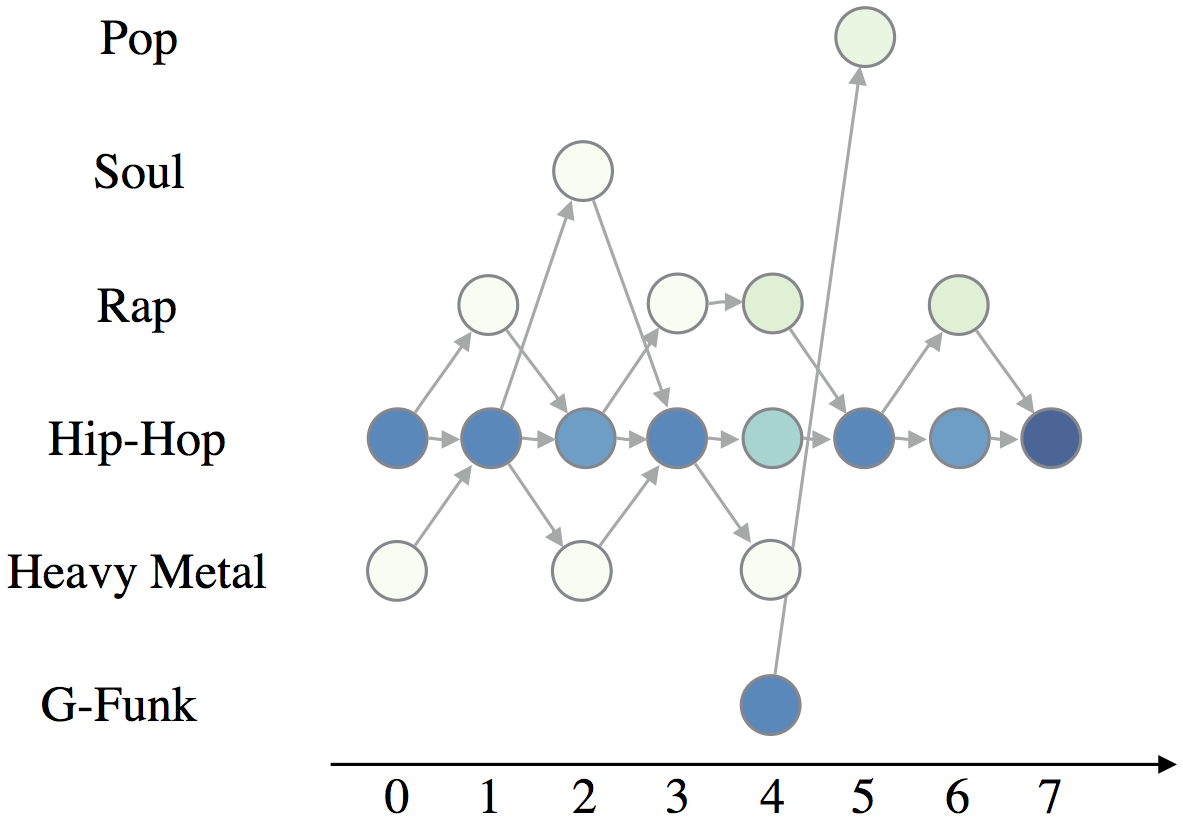}}
    \end{tabular}
\else
    \begin{tabular}{cc}
    \subfloat[Metal component. While the genre diversity is a bit higher than in the Jazz example (b), all genres belong to the same meta-genre: Metal.\label{Fig12}]{\includegraphics[width=\columnwidth]{Fig12.png}} &
    \subfloat[Jazz component. We clearly see that Jazz playlists are very pure and do not mix with other genres. The same phenomenon also exists for other music niches such as Classical music.\label{Fig13}]{\includegraphics[width=\columnwidth]{Fig13.png}} \\
    \subfloat[Rock - Alternative component. The diversity of the genres is much higher than in previous examples but stays coherent.\label{Fig15}]{\includegraphics[width=\columnwidth]{Fig15.png}} &
    \subfloat[Rap / Hip-hop component. Heavy Metal nodes are not outliers but are indeed connected to Hip-Hop with songs from famous artists such as Korn, Limp Bizkit, Public Enemy, etc. \label{Fig14}]{\includegraphics[width=\columnwidth]{Fig14.png}}
    \end{tabular}
\fi
\caption{{\changes{\bf Average activated component.} For a clearer visualization, we have discarded nodes with a low probability of appearance. The horizontal axis represent the layers: the position in the playlist (limited to the 8 first layers). The nodes are colored from light to dark according to their weights, i.e. their likelihood of appearance (larger is darker).}}
\label{Fig:music}
\end{center}
\end{figure*}

\subsubsection{The graph} A graph, $G$, is constructed by doing the union of all the playlists in the dataset. As a consequence, an edge between two songs only exists if at least one playlist contains this particular sequence of songs. This graph thus encodes song ``affinity'' together with their \emph{causal relationships}. The number of nodes of $G$ is over $159,000$. Every edge in the graph has been created from an actual human-made audio playlist. It is thus perfectly suited for building new playlists following human tastes.

\subsubsection{The signal and mask} As a signal on the graph, we use the likelihood that a song is in a particular place on a playlist. To compute it, we count how many times a particular song has been placed at a particular position for all playlists. For example, take a song A which has been placed 3 times in the first position of a playlist and 5 times in the third position. Thus, in this example, the vector on the node A has 2 non-zero values $(3,0,5,0,\cdots)$. The number of entries of each vector is equal to the number of songs in the longest playlist. Similarly to the previous application, we normalize the signal by z-score, giving a likelihood to be at a given position in a playlist. As a consequence, song with positions evenly distributed in playlists do not reach the threshold. Only songs {\changes appearing at a limited number of positions in playlists e.g only in the first and third position in the previous example}, have non-zero binary values for these positions. Note that having well-defined locations for songs is important as we use their ordering in playlists as causal relationships between them. Setting $\mu=0.1\sigma$ proves to be a reasonable choice of threshold as the extracted components have an average width of around $9$ steps: it is close to the average length of a handcrafted playlist in the dataset.

\subsubsection{Playlist recommender system} 
We propose an algorithm to generate playlists based on music ``moods'' using average activated components obtained by $k$-means clustering. As an outcome of our method we show that different music moods are associated, in a totally unsupervised manner, to the clusters. Moods are ``Electroish'', ``Metallic'', ``Rocky'' (see Fig.~\ref{Fig12} for example) and can be viewed as a meta-genre of music regrouping related music genres such as Rock, Indie, Alternative, etc. We extract activated components of songs from the causal multilayer graph by thresholding the normalized popularity vector (keeping the largest peaks of appearance of songs in playlists). Each activated component is a group of songs fitting nicely together and respecting a precise order within the playlists. These activated components are then clustered together using $k$-means clustering. {\changes The number of clusters $k=30$ is here naturally given by the number of dominant genres present in the dataset. This is a natural choice as we expect playlists to be classified by genre. In addition, music genre often proves to be one of the most important criteria when creating a mix: in the dataset, half of the distribution of the playlist categories are labeled with a music genre. This choice of $k$ is validated by the results which are indeed meaningful in term of genres.
}

Each cluster can be labeled with a mood \emph{a posteriori} by analyzing music genres present within the clusters. Once the mood has been chosen by a user, the algorithm selects the average activated component associated to the mood. To generate a playlist, a seed song, the first of the playlist, is selected in the first layer of the component. In its simplest form, the selection is done at random. However, many criteria such as user history, ratings or time since last played, can be used to select a starting point. Then, the rest of the playlist is constructed by doing a random walk on the \emph{causal} edges of the component. The familiarity versus discovery ratio can be tuned by modifying the edge weights according to the popularity of each node. The random walker would have more or less chances to reach a popular song depending on the user's will.

A good playlist should alternate between familiarity, discovery and smoothness in transitions between songs \cite{mcfee2011natural}. An average activated component is a coherent weighted subgraph of songs, where each node and edge are weighted by their likelihood of appearance at that particular position. The most popular songs, appearing in many activated components, have a large weight, filling the contract for the familiarity part. Less popular songs will also be clustered together giving choice for the discovery part. Finally, the smoothness of transitions is guaranteed by the graph $G$: each song to song transition has been created by a human and is appealing to at least one of them.

Keeping the original ordering of songs in a playlist (successive positions of several songs, not just 2) has been shown to be crucial when designing recommender systems as shown in~\cite{mcfee2011natural} and~\cite{Bonnin_2014}. Following the causal edges of an average activated component takes into account the ordering of several successive songs with their positions within playlists, unlike a random walk on the graph $G$ which considers only one to one coupling between songs.

\begin{figure}[!ht]
\begin{center}
\ifdraft
    \includegraphics[width=0.6\columnwidth]{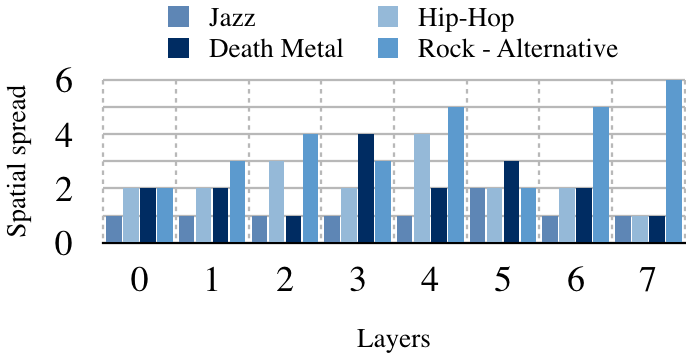}
\else
    \includegraphics[width=0.9\columnwidth]{Fig16.png}
\fi
\caption{ {\bf The spatial spread of the different average activated components presented in Fig.~\ref{Fig12}, Fig.~\ref{Fig13}, Fig.~\ref{Fig14}, Fig.~\ref{Fig15}.} The spatial spread is the number of genres per layer of the average activated component. As one could expect, popular genres such as Rock and Alternative have a bigger spatial spread and are easily mixed with other genres.
}\label{Fig16}
\end{center}
\end{figure}

\subsubsection{Visualization} The activated components can also be used for exploring and visualizing the dataset. Since each activated component has a large spatial spread, we group songs on each layer by genre as it drastically reduces the dimensionality and exhibits interesting insights on how users create playlists. Note that genres are here to validate the methodology and have not been used to cluster the activated components together. Like in previous applications the method is completely unsupervised. The results are shown in Fig.\ref{Fig12}, Fig.\ref{Fig13}, Fig.\ref{Fig14}, Fig.\ref{Fig15}.

While the dataset is biased towards Rock, Alternative and Indie (more than 40\% of all the songs), the clustering still achieves to extract relatively pure patterns of related genres, or music ``moods'', as it is shown in Fig.~\ref{Fig12} and Fig.~\ref{Fig13}. As music experts could have expected, songs of popular genres are more volatile: they can easily be mixed with other genres and have a higher spatial spread, see Fig.~\ref{Fig15} and Fig.~\ref{Fig16}. On the contrary, songs of ``connoisseur genres'' such as Metal, Jazz, Hip-Hop or Classical stay clustered in their universe.

\section{Conclusion}

From a general point of view, we have presented a new framework to extract and analyze sparse repeated patterns created by dynamical processes on graphs. Our approach is based on the causal multilayer graph, a novel multilayer graph structure that encodes the propagation or spreading of events across time.

The construction of the causal multilayer graph and the extraction of dynamic activation components are computationally efficient and fully leverage today's multicore architecture. By applying our framework in three different real-world applications, we have demonstrated that clustering similar patterns of activity and analyzing average activation components reveals new insights on the underlying causal processes.

In addition to the applications presented here, our method can also be applied to problems actually modeled as temporal networks, allowing a different approach and an additional degree of model complexity. More generally, we believe that our model shows great promise for a wider range of problems such as the spreading of epidemic outbreaks, social network activity, brain EEG recordings or any type of sensor networks. Applications where time series have been recorded on the vertices of a network are numerous, present in many fields of science such as engineering, social, biological, physical or computational science and keep increasing with the actual data deluge.

% % \newpage
\section*{Acknowledgements}
This work was partially funded by SNF grant number 200021\_154350 1.

\appendix

\subsection{Efficient construction of the causal multilayer graph of activity}
\label{sec:efficient_construction}

In this section, we describe the different steps to construct the CMG of activity, in a fast manner, directly from the spatial graph, $G$, and the mask, $\mathbf{M}$. For clarity, several technical details aimed at optimizing further the implementation, but which are not crucial for the understanding, are given in the next section. 

\begin{figure}[h]
\begin{center}
\ifdraft
    \includegraphics[width=0.7\columnwidth]{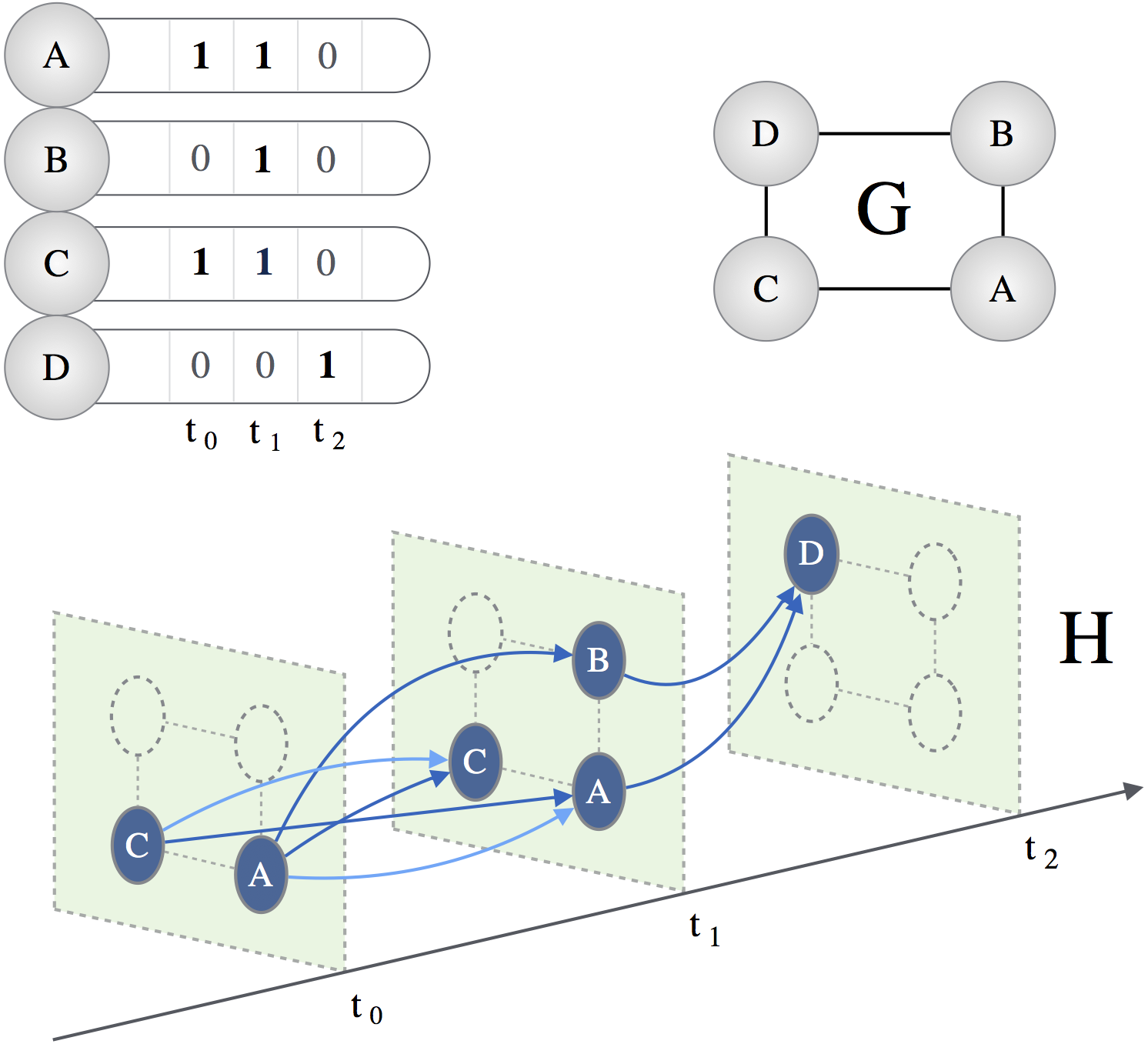}
\else
    \includegraphics[width=1\columnwidth]{Fig5.png}
\fi
\caption{
{\bf Actual implementation of the CMG of activity.} The CMG, $H$, is directly constructed from the activated entries of the binary mask (top left) and the original graph, $G$ (top right). The construction of $K$ is not needed.
}\label{Fig5}
\end{center}
\end{figure}

We first combine the data into one object. Each binary activation vector, $M(i,\cdot)$, is stored as a property (label) on node $i$ of $G$, creating a property graph which is still denoted $G$.
The causal multilayer graph of activity $H$ is created in a series of steps illustrated in Fig.~\ref{Fig5} and detailed as follows:
\begin{enumerate}
\item Iterating over each edge, $e\in E_G$, of $G$ linking a source node, $i$, and a destination node $j$, the algorithm first reads the vectors $M(i,\cdot)$ and $M(j,\cdot)$. An inter-layer connection is created between layer $t$ and $t+1$ whenever $i_t$ and $j_{t+1}$ are activated (since we already know that $i$ and $j$ are spatial neighbors) i.e. $M(i,t)=M(j,t+1)=1$. That is to say, an edge exists in $\Omega_x$ whenever $M(i,t)\ \&\  M(j,t+1)=1$ where $\&$ is the logical And. We introduce a new vector $u_e$ of size $T-1$ associated to the edge $e$, for all $t\in [0,\cdots,T-2]$:
\begin{equation}\label{eq:edgebuilding}
u_e(t)= M(i,t)\ \&\  M(j,t+1).
\end{equation}
The value $u_e(t)$ encodes the existence (1) or absence (0) of an inter-layer edge between vertices $i_t$ and $j_{t+1}$. The vector $u_e$ is stored as a property of edge $e$.
\item Since we are also interested in self activation of vertices across time (the set $\Omega_{xs}$) we compute an additional vector $u_{{\rm self},i}$ for each \emph{vertex} $i$ of $G$,
\begin{equation}\label{eq:selfedgebuilding}
u_{{\rm self},i}(t)= M(i,t)\ \&\  M(i,t+1),
\end{equation}
for all $t\in [0,\cdots,T-2]$. 
The vector $u_{{\rm self},i}$ is stored as a property of node $i$.
\item The construction of the graph $H$ is then done by reading the collection of edge vectors, $\{u_e\}_e$, node vectors, $\{u_{{\rm self},i}\}_i$, and adding edges between successive time layers when ones are encountered.
\end{enumerate}

The complexity of this process is linear in the number of edges and vertices of $G$. It is also linear in the number of time-steps. The scalability of our method comes from the properties of $H$. Layers are time-ordered, allowing connections between layers to be encoded as binary vectors. The creation of edges only depends on the state of pairs of nodes. Therefore, it relies on local values, which allows for, an efficient parallel implementation. The main task consists in handling the properties (vectors) of triplets, $\{i, e, j\}$, where $e \in E_G$ is the spatial edge connecting source node, $i$, and destination node, $j$. This is sufficient to create all causal edges between $i$ and $j$ ($i\neq j$). 
This interesting property is particularly suited to large scale graph analytics frameworks such as \href{https://dato.com/products/create/index.html}{GraphLab Create} \cite{Low_2012} or Apache GraphX~\cite{xin2013graphx}, which have dedicated processes for applying functions to all triplets in parallel. Thus, our implementation gracefully scales with the number of cores and is much faster than a sequential naive implementation. To our knowledge, no other implementation of multilayer graphs matches the speed and the scalability of the method proposed here.

\subsection{Implementation details}
This section describes additional implementational tricks for a better computer efficiency of the method. The reader interested in the general method more than its implementation may skip it.

The first step of the algorithm reads both activation mask vectors $M(i,t)$, $M(j,t+1)$ from $i$ and $j$, respectively, as arbitrary precision integers \cite{MacLaren_1970}. An arbitrary precision integer can store any integer number (limited by available memory), and can be seen as a list of ``standard'' $32$ bit integers with a common interface. GraphLab Create only allows the storage of basic datatypes as properties of nodes and edges such as integer, double, bool, and string (at the time of the writing). We had to transform the rows of our activation mask, $M$, to bitstring to be able to store them on vertices and edges. Any bit compression algorithm can be used to reduce storage space. The arbitrary precision integer stored as a string offers a compression ratio of more than 3 over the raw bitstring.

\begin{figure}[h!]
\begin{center}
\ifdraft
    \includegraphics[width=0.7\columnwidth]{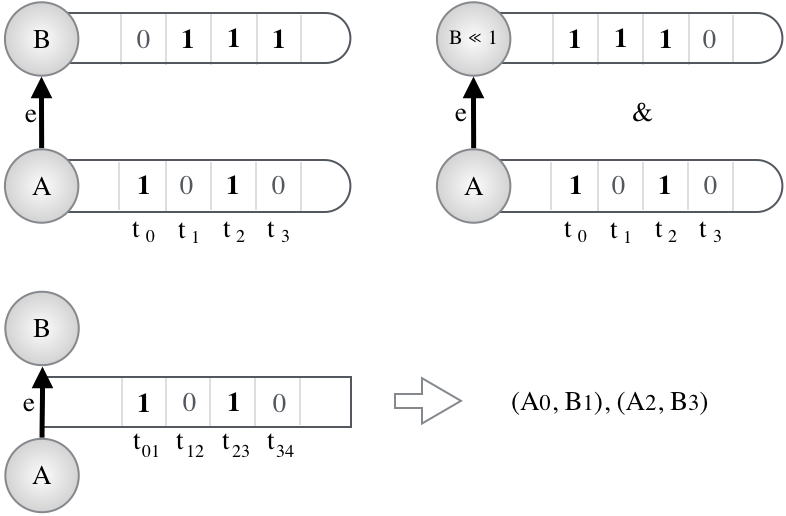}
\else
    \includegraphics[width=1\columnwidth]{Fig6.png}
\fi
\caption{{\bf Creation of causal edges from a triplet $(A, e, B)$.} The binary mask vector of the destination node, $B$, is shifted (in this illustration left-shifted) to align source layer, $t$, and destination layer $t+1$. Then, a logical And is performed between the two vectors (top right picture). On the bottom figure, the result $u$, stored in $e$, is read to create the edges (and nodes) of $H$.%
}\label{Fig6} 
\end{center}
\end{figure}

In the second step of the algorithm, the vector $u_e(t)$ is created by performing a logical And ($\&$) between $M(i,t)$ and $M(j,t+1)$. It amounts to performing a logical And between two vectors: $M(i,\cdot)$ and the left-shift version of $M(j,\cdot)$ (hence involving a logical And and a bit shift). The last (least significant) bit in this operation is dropped ($u_e$ is of size $T-1$) as it would correspond to a link between layer $T-1$ and $T$, the latter of which does not exist. Once the vector $u_e(t)$ is created, all the ones have to be found to create the causal edges. For each $1$, its position in the vector gives the time $t$ and allows the creation of two pairs, $(i, t)$ and $(j, t+1)$, a causal edge, which is then added in the causal multilayer graph of activity $H$.
Instead of looping through all the bits of $u_e(t)$ and checking for a $1$ at each position, we have implemented another strategy that jumps from $1$ to $1$ in $u_e(t)$ and gives the position of the layer number $t$. This optimization is better than the classical For loop when $u_e(t)$ is sparse. The details are given in Algorithm~\ref{algo}. We invite the interested reader to refer to~\cite{anderson2005} for more information on this low level bit manipulation trick. The Fig.~\ref{Fig6} illustrates the algorithm formally defined in Algorithm~\ref{algo}.

\begin{algorithm}[h!]
    \caption{ {\bf Creation of the causal multilayer graph $H$ on little-endian systems.} The least significant bit is on the right.}
    
    \newcommand*\BitAnd{\mathrel{\&}}
    \newcommand*\BitOr{\mathrel{|}}
    \newcommand*\ShiftLeft{\ll}
    \newcommand*\ShiftRight{\gg}
    \newcommand*\BitNeg{\ensuremath{\mathord{\sim}}}
    
    \SetKw{Parallel}{Parallel}
    \SetKwFunction{AddEdge}{AddEdge}

    \KwIn{Graph $G$ having the binary activation vectors stored on its nodes}
    \KwOut{Graph $H$}
    \BlankLine

    $H \longleftarrow$ create empty directed graph\;
    \Parallel{} \ForEach{triplet $(src, \, e, \, dst) \in G$} {
        $m_s \longleftarrow $ read src binary vector $m$ as arbitrary-sized integer\;
        $m_d \longleftarrow $ read dst binary vector $m$ as arbitrary-sized integer\;
        $u \longleftarrow m_s \BitAnd (m_d \ShiftRight 1)$\;
        \BlankLine

        \tcp{Find all activated causal edges}
        \While{u is not $0$} {
            \tcp{extract least significant bit on a 2s complement machine}
            $index \longleftarrow u \BitAnd -u$\;
            $u \longleftarrow u \oplus index$ \tcp*[h]{toggle the bit off}\;

            \BlankLine

            \tcp{Get activated layer number}
            $layer \longleftarrow -1$\;
            \While{index is not $0$} {
                $index \longleftarrow index \ShiftRight 1$\;
                $layer \longleftarrow layer + 1$\;
            }
            \AddEdge{H, (src, layer), (dst, layer+1)}\;
        }
    }
\label{algo} 
\end{algorithm}

\subsection{The generalized causal multilayer graph of activity}
\label{sec:generalized}

{\changes In Sec.~\ref{sec:act_cmg}, we introduced the mathematical foundations of the causal multilayer graph of activity. Here, we generalize this model to account for dynamic spatial graphs, where edges or nodes are allowed to appear or disappear across layers. By encoding the position of nodes and edges as additional vectors on the nodes and edges of the generalized spatial graph $G$, the generalized CMG of activity can be constructed very efficiently using a variation of the algorithm introduced in Appendix~\ref{sec:efficient_construction}.

Let us assume we have a set $\{G_t\}_t$ of $T$ graphs, one for each time-step $t\in[0,\cdots,T-1]$. The number of vertices and edges is allowed to change from graph to graph.
In addition, a mask $M$ that associates a value (0 or 1) to each vertex of the set $\{G_t\}_t$ is given. In this case, it may not be a matrix. The mask can also be computed from a signal $S$ on the vertices of the collection of graphs $\{G_t\}_t$.

We first create the generalized spatial graph $G$, which concatenates all $\{G_t\}_t$. A vertex $i$ belongs to $V_G$ if there exist some layer $t$ such that $i_t\in V_{G_t}$. Similarly, an edge $e\in E_G$ connecting vertex $i$ and $j$ of $G$ exists if for some $t$ there is an edge $e_t\in G_t$ connecting $i_t$ and $j_t$. Taking the definition of Eq.~\eqref{eq:CMGset}, the expression of the weights in Eq.~\eqref{eq:tensordef} is $w_{ij}(t(t+1))=W_{i,j}(t)$ for $i\neq j$, where $W_{i,j}(t)$ is the weight of the edge between $i_t$ and $j_t$; For $i=j$, $w_{ii}(t(t+1))=1$ and zero for the rest.

The second step associates vectors to each vertex and edge of $G$. Similarly to Sec.~\ref{sec:act_cmg}, a vector $M(i,\cdot)$ of length $T$ is associated to each vertex $i$. Additional vectors, i.e values of the signal $S_i$ can also be stored on each vertex. In the case where vertex $i$ is not present in some graph $G_t$ the entry $M(i,t)$ exists and is set to zero. It is equivalent to adding an extraneous inactive and unconnected vertex $i$ at layer $t$. We now introduce an additional mask: the edge mask $M_E$. It associates a binary vector of length $T$ to each edge of $G$ i.e.: for each edge $e$ of $G$ between node $i$ and $j$, $M_E(e,t)=1$ if there is an edge between $i_t$ and $j_t$, zero otherwise. Notice that edge weights can also be stored as vectors on each edge, similarly to what is done to the signal $S$ on vertices. For the sake of simplicity, we assume the graphs to be unweighted here. 

For an efficient construction, we modify~\eqref{eq:edgebuilding} to account for the existence of edges between layers. It becomes:
\begin{equation}\label{eq:edgebuildinggeneral}
u_e(t)= M(i,t)\ \&\  M(j,t+1)\ \&\ M_E(t).
\end{equation}

In the case of self-edges, the process given by Eq.\eqref{eq:selfedgebuilding} is unchanged: two vertices will be connected if they exist in two successive layers and are active.

Fig.~\ref{Fig19} explains the construction of the generalized CMG of activity using the different vectors. Similarly to the standard model, for each edge $e$ the destination vertex mask is shifted before performing a logical And with the source vector. The only difference is the introduction of the edge mask $M_e$ which is logically ``Anded''  between the source and shifted destination masks. We use the same process at the vertex level to account for self-edges (connections between $i_t$ and $i_{t+1}$).

\begin{figure}[h!]
\begin{center}
\ifdraft
    \includegraphics[width=0.7\columnwidth]{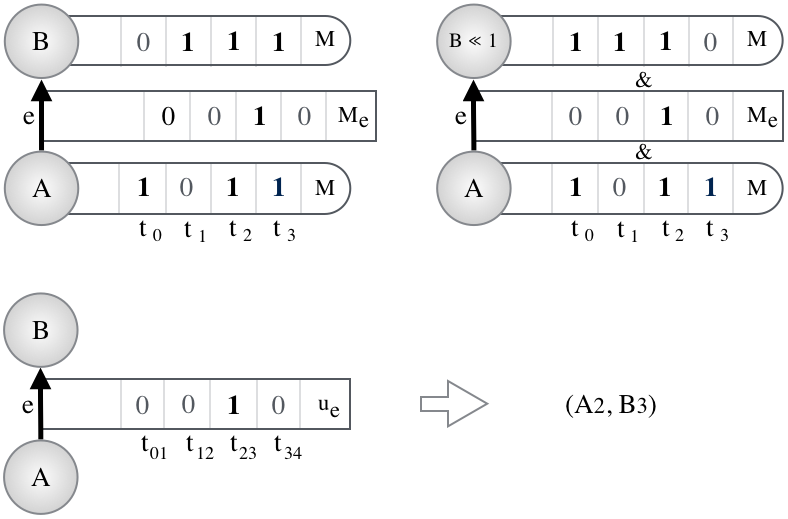}
\else
    \includegraphics[width=1.0\columnwidth]{Fig19.png}
\fi
\caption{{\changes{\bf Creation of causal edges in the generalized CMG of activity.} Top left: a source node $A$ and a destination node $B$ connected by edge $e$, together with their respective mask. Top right: the binary mask vector of $B$, is shifted (in this illustration left-shifted). Then, a logical And is performed between the three vectors. On the bottom figure, the result $u$, stored in $e$, is read to create the edges (and nodes) of $H$. In this example, the pair $(A_{t_0}, B_{t_1})$ is disconnected because the first entry of $M_E$ is $0$. Only one edge between $A_2=A_{t_2}$ and $B_3=B_{t_3}$ is created.%
}}\label{Fig19}
\end{center}
\end{figure}

The rest of the construction leading to $H$ is similar to the simplified case. Once $H$ has been constructed, the creation of feature vectors and the clustering are identical.}

\newpage
% \bibliography{biblio.bib}

\end{document}